\documentclass{ar2e}%%%Put [range] option here for references citations to
\usepackage{graphicx}
\usepackage{url}
\usepackage{color}
\usepackage{amsmath,amssymb}
\usepackage[hyperindex,breaklinks]{hyperref}

\def\ode{\Omega_{\rm DE}}

\usepackage{ARAstroBib}

\begin{document}

\input INCLUDE/epsf.tex    %<-If you need EPS figures to be
                   %  called in {figure} environment for PC
\input INCLUDE/epsf.def   %<-If you need EPS figures to be
                   %  called in {figure} environment for Macintosh

\input INCLUDE/psfig.sty

\def\nat{{ Nature }}
\def\aap{{ Astron. \& Astrophys. }}
\def\aj{{ Astron.~J. }}
\def\apj{{ Astrophys.~J. }}
\def\araa{{ Ann. Rev. Astron. Astrophys. }}
\def\apjl{{ Astrophys.~J.~Letters }}
\def\apjs{{ Astrophys.~J.~Suppl. }}
\def\apss{{ Astrophys.~Space~Sci. }}
\def\icarus{{ Icarus }}
\def\mnras{{ MNRAS }}
\def\pasp{{ Pub. Astron. Soc. Pacific }}
\def\planss{{ Plan. Space Sci. }}
\def\physrep{{ Phys. Rep.}}
\def\bain{{ Bull.~Astron.~Inst.~Netherlands }}
\def\lesssim{\mathrel{\hbox{\rlap{\hbox{\lower4pt\hbox{$\sim$}}}\hbox{$<$
}}}}

\def\cc{\mbox{cm$^{-3}$}}
\def\tauv{\mbox{$\tau_V$}}
\def\av{\mbox{$A_V$}}
\def\ra{\mbox{$\rightarrow$}}
%                                                  units
\def\d{^\circ}
\def\h{^{\rm h}}
\def\mi{^{\rm m}}
\def\s{^{\rm s}}
\def\mum{\ts \mu{\rm m}}
\def\mm{\ts {\rm mm}}
\def\cm{\ts {\rm cm}}
\def\percm{\ts {\rm cm}^{-1}}
\def\m{\ts {\rm m}}
\def\kms{\rm{\, km \, s^{-1}}}
\def\K{\ts {\rm K}}
\def\Kkms{\ts {\rm K\ts km\ts s^{-1}}}
\def\kHz{\ts {\rm kHz}}
\def\MHz{\ts {\rm MHz}}
\def\GHz{\ts {\rm GHz}}
\def\pc{\ts {\rm pc}}
\def\kpc{\ts {\rm kpc}}
\def\Mpc{\ts {\rm Mpc}}
\def\cmsq{\ts {\rm cm^2}}
\def\pcsq{\ts {\rm pc^2}}
\def\dsq{\ts {\rm deg^2}}
\def\debye{\ts10^{-18}\ts {\rm esu}\ts {\rm cm}}
\def\swash2o{$1_{10} - 1_{01}$}             

%                                                  journals
%                                                  mathe
\let\ap=\approx
\let\ts=\thinspace

\def\an{{ Astronomische Nachrichten }}
\def\sci{{ Science }}
\def\prl{{ Phys. Rev. Lett. }}
\def\zfa{{ Zeitschrift fur Astrophysik }}
\def\ba{{ Baltic Astronomy }}
\def\rmp{{ Rev. Mod. Phys. }}
\def\rpp{{ Rep. Prog. Phys. }} 
\def\pasj{{ Pub. Astron. Soc. Japan }}
\def\pr{{ Phys. Rev. }}
\def\grg{{ Gen. Rel. Grav. }}
\def\sitz{{ Sitzungsber. K. Akad. }}

\jname{Annu. Rev. Astron. Astrophys.}
\jyear{2008}
\jvol{46}
\ARinfo{XXX}

\title{Dark Energy and the Accelerating Universe}

\markboth{Frieman, Turner \& Huterer}{Dark Energy}

\author{Joshua A. Frieman
\affiliation{Center for Particle Astrophysics, Fermi National Accelerator Laboratory, P. O. Box 500, Batavia, IL 60510; \\
Kavli Institute for 
Cosmological Physics, The University of Chicago, 5640 S. Ellis Ave., 
Chicago, IL 60637; email: frieman@fnal.gov}
Michael S. Turner
\affiliation{
Kavli Institute for Cosmological 
Physics, The University of Chicago, 5640 S. Ellis Ave., Chicago, IL 60637; 
email: mturner@uchicago.edu \\
}
Dragan Huterer
\affiliation{
Department of Physics, University of Michigan, 450 Church St., Ann Arbor, MI, 48109; 
email: huterer@umich.edu}
}

\begin{keywords}
cosmology, cosmological constant, supernovae, galaxy clusters, large-scale 
structure, weak gravitational lensing
\end{keywords}

\begin{abstract}

%% mst 10.29.07
%% mst 01.18.08
%% mst 01.29.08

The discovery ten years ago that the expansion of the Universe is accelerating put in place the last major building block of the present cosmological model, in which the Universe is composed of 4\% baryons, 20\% dark matter, and 76\% dark energy. At the same time, it posed one of the most profound mysteries in all of science, with deep connections to both astrophysics and particle physics. Cosmic acceleration could arise from the repulsive gravity of dark energy -- for example, the quantum energy of the vacuum -- or it may signal that General Relativity breaks down on cosmological scales and must be replaced. We review the present observational evidence for cosmic acceleration and what it has revealed about dark energy, discuss the various theoretical ideas that have been proposed to explain acceleration, and describe the key observational probes that will shed light on this enigma in the coming years.

\end{abstract}

\maketitle

\section{INTRODUCTION}
\label{intro}
%% Text for Section 1 of FTH ARAA

In 1998, two teams studying distant Type Ia supernovae
presented independent evidence that the expansion of the Universe is speeding
up \citep{riess98,perlmutter99}.    Since Hubble, cosmologists
had been trying to measure the slowing of the expansion due to gravity; so
expected was slow-down that the parameter used to quantify the second
derivative of the expansion, $q_0$, was called the deceleration parameter
\citep{Sandage62}. The discovery of cosmic acceleration 
is arguably one of the most
important developments in modern cosmology.

The ready acceptance of the supernova results was not a foregone conclusion. 
The cosmological constant, the simplest explanation of accelerated expansion,
had a checkered history, having been invoked and subsequently withdrawn
several times before. This time, however, 
subsequent observations, including
more detailed studies of supernovae and independent evidence from clusters of
galaxies, large-scale structure, and the cosmic microwave background (CMB), 
confirmed and firmly established this remarkable finding.

The physical origin of cosmic acceleration remains a deep mystery.  According to General
Relativity (GR), if the Universe is filled with ordinary matter or radiation,
the two known constituents of the Universe, gravity should lead to a slowing
of the expansion.  Since the expansion is speeding up, we are faced with two
possibilities, either of which would have profound implications for our
understanding of the cosmos and of the laws of physics. The first is that 75\%
of the energy density of the Universe exists in a new form with
large negative pressure, called dark energy.
The other possibility is that General Relativity breaks down on cosmological
scales and must be replaced with a more complete theory of gravity.

Through a tangled history, dark energy is tied to Einstein's cosmological
constant, $\Lambda$.  Einstein introduced $\Lambda$ into the field equations of General Relativity
in order to produce a static, finite cosmological model
\citep{einstein17}. With the discovery of the expansion of the Universe, the
rationale for the cosmological constant evaporated.  Fifty years later,
\citet{zeldovich68} realized that $\Lambda$, mathematically equivalent to the
stress-energy of empty space---the vacuum---cannot simply be dismissed. 
In quantum field theory, the vacuum state is filled with
virtual particles, and their effects have been measured in the shifts of atomic
lines
and in particle masses. 
However, estimates for the energy density associated with the quantum
vacuum are at least 60 orders of magnitude too large and in
some cases infinite, a major embarrassment known as the cosmological constant
problem \citep{WeinbergRMP}.

Despite the troubled history of $\Lambda$, the observational evidence for 
cosmic acceleration was quickly embraced by cosmologists, 
because it provided the missing element needed to complete the current cosmological model.  
In this model, the Universe is spatially flat and accelerating;
composed of baryons, dark matter, and dark energy; underwent a hot, dense, early
phase of expansion that produced the light elements via big bang
nucleosynthesis and the CMB radiation; and experienced a much earlier epoch of
accelerated expansion, known as inflation, which produced density perturbations
from quantum fluctuations, leaving an imprint on the CMB anisotropy and
leading by gravitational instability to the formation of large-scale
structure.  

The current cosmological model also raises deep issues, from the origin of
the expansion itself and the nature of dark matter to the genesis of baryons and
the cause of accelerated expansion.  Of all these, the mystery of cosmic
acceleration may be the richest, with broad connections to other important
questions in cosmology and in particle physics.  For example, the destiny
of the Universe is tied to understanding dark energy; primordial inflation
also involves accelerated expansion and its cause may be related; dark matter
and dark energy could be linked; cosmic acceleration could provide a key to
finding a successor to Einstein's theory of gravity; the smallness of the
energy density of the quantum vacuum might reveal something about
supersymmetry or even superstring theory; and the cause of cosmic acceleration
could give rise to new long-range forces or be related to the smallness of
neutrino masses.

This review is organized into three parts.  The first part is devoted to {\em
Context:} in \S 2 we briefly review the Friedmann-Robertson-Walker (FRW) cosmology, 
the framework for understanding how observational probes of dark energy work.
\S 3 provides the historical context, from Einstein's introduction of the 
cosmological constant to 
the supernova discovery.
Part Two covers {\em Current Status:} in \S 4, we review the web of
observational evidence 
that firmly establishes accelerated expansion.
\S 5 summarizes current 
theoretical approaches to accelerated expansion and dark energy, including
discussion of the cosmological constant problem, models of dark energy, and
modified gravity, while \S 6 focuses on different phenomenological
descriptions of dark energy and their relative merits.  Part Three addresses
{\em The Future:}
\S 7 discusses the observational techniques that will be used to probe dark energy, 
primarily supernovae, weak lensing, large-scale structure, and clusters.
In \S 8, we discuss specific projects aimed at constraining
dark energy planned for the next fifteen years which have the potential to provide insights into the origin of cosmic acceleration.  The
connection between the future of the Universe and dark energy is the topic of
\S 9.  We summarize in \S 10, framing the two big questions about cosmic
acceleration where progress should be made in the next fifteen years -- 
Is dark energy something other than vacuum energy? Does General Relativity self-consistently describe cosmic acceleration? -- and discussing what we believe are the most important open issues.

Our goal is to broadly review cosmic acceleration for the astronomy community.  
A number of useful reviews target different aspects of the subject, including:   
theory \citep{Copeland_review,Padmanabhan_review}; cosmology \citep{Peebles_Ratra_03};  the physics of cosmic acceleration \citep{Uzan_06}; probes of dark energy 
\citep{Hut_Tur_00}; dark energy reconstruction \citep{Sahni_review}; dynamics
of dark energy models \citep{Linder_review}; the
cosmological constant \citep{carroll92,Carroll_LivRevRel}, and the 
cosmological constant problem \citep{WeinbergRMP}.

\section{BASIC COSMOLOGY}
\label{cosmology}
In this section, we provide a brief review of the elements of the
FRW cosmological model. This model provides the
context for interpreting the observational evidence for cosmic
acceleration as well as the framework for understanding how
cosmological probes in the future will help uncover the cause of acceleration
by determining the history of the cosmic expansion with greater precision.
For further details on basic cosmology, see, e.g., the textbooks of
\citet{Dodelson_book,Kolb_Turner,Peacock_book}, and \citet{Peebles_book}. 
Note that we follow the standard practice of using units in which the speed of
light $c=1$.

\subsection{Friedmann-Robertson-Walker cosmology}
From the large-scale distribution of galaxies and the near-uniformity of the
CMB temperature, we have good evidence that the Universe is nearly 
homogeneous and
isotropic.  Under this assumption, the spacetime metric can be written in the FRW form,
\begin{equation} 
ds^2 = dt^2 - a^2(t) \left[ dr^2/(1-kr^2) +r^2 d\theta^2 +r^2 \sin^2 \theta d\phi^2 \right]  ,
\label{eq:metric}
\end{equation}
where $r, \theta, \phi$ are comoving spatial coordinates, $t$ is time,  and 
the expansion is described by the cosmic scale factor, $a(t)$ (by 
convention, $a=1$ today). 
The quantity $k$ is the curvature of 3-dimensional space: $k = 0$ corresponds
to a spatially flat, Euclidean Universe, $k>0$ to positive curvature
(3-sphere), and $k<0$ to negative curvature (saddle).

The wavelengths $\lambda$ of photons moving through the
Universe scale with $a(t)$, and the redshift of light emitted from a distant
source at time $t_{\rm em}$,  
$1 + z = \lambda_{\rm obs}/\lambda_{\rm em} =1/a(t_{\rm em})$,
directly reveals the relative size of the Universe at that time.
This means that time intervals are related to redshift intervals by
$dt = -dz/H(z)(1+z)$, where $H\equiv \dot a/a$ is the Hubble parameter, and an
overdot denotes a time derivative. The present value of the Hubble parameter
is conventionally expressed as $H_0 = 100 ~h$ km/sec/Mpc, where $h\approx 0.7$ is the 
dimensionless Hubble parameter. Here and below, a
subscript ``0'' on a parameter denotes its value at the present epoch.

The key equations of cosmology are the Friedmann equations, the 
field equations of GR applied to the FRW metric, 
\begin{eqnarray}
H^2 = \left(\frac{\dot a}{a}\right)^2 & = & {8\pi G\rho\over 3} -{k\over a^2}
\ \  + {\Lambda\over 3}  
\label{eq:feq}
\\
\frac{\ddot a}{a}               & = & -{4\pi G\over 3}\,(\rho + 3p) \ \ 
 + {\Lambda\over 3} 
%\label{eq:feq2}
\end{eqnarray}
where $\rho$ is the total energy density of the Universe (sum of matter,
radiation, dark energy), and $p$ is the total pressure (sum of pressures of
each component).  For historical reasons we display the cosmological constant
$\Lambda$ here; hereafter, we shall always represent it as vacuum energy and
subsume it into the density and pressure terms; the correspondence is:
$\Lambda = 8\pi G\rho_{\rm VAC}=-8\pi G p_{\rm VAC}$.

For each component, the conservation of energy is expressed by $d(a^3\rho_i )
= -p_i da^3$, the expanding Universe analogue of the first law of 
thermodynamics, $dE=-pdV$.  Thus, the
evolution of energy density is controlled by the ratio of the pressure to the
energy density, the equation-of-state parameter, $w_i \equiv p_i/\rho_i$. 
\footnote{A perfect fluid is fully characterized by its isotropic pressure $p$ and energy density $\rho$, where $p$ is a function of density and other state variables (e.g., temperature). The equation-of-state parameter $w=p / \rho$ determines the evolution of the energy density $\rho$; e.g., $\rho \propto V^{1+w} $ for constant $w$, where $V$ is the volume occupied by the fluid.  Vacuum energy or a homogeneous scalar field are spatially uniform and they too can be fully characterized by $w$. The evolution of an inhomogeneous, imperfect fluid is in general complicated and not fully 
described by $w$. Nonetheless, in the FRW cosmology, spatial homogeneity and 
isotropy require the stress-energy to take the 
perfect fluid form; thus, 
$w$ determines the evolution of the energy density.} 
For the general case, this ratio varies with time, and the
evolution of the energy density in a given component is given by
\begin{equation}
\rho_i \propto \exp \left[ 3 \int_0^z [1+w_i(z^\prime )]d\ln (1+z^\prime ) \right] ~.
\label{eq:rho_scaling}
\end{equation}
In the case of constant $w_i$, 
\begin{equation}
w_i \equiv {p_i\over \rho_i}= {\rm constant}~,~~  \rho_i  \propto   (1+z)^{3(1+w_i)}~.
\label{eq:rhow}
\end{equation}
For non-relativistic matter, which includes both dark matter and baryons, $w_{\rm M}
= 0$ to very good approximation, and $\rho_{\rm M}
\propto (1+z)^{3}$; for radiation, i.e., relativistic particles, $w_{\rm R}=1/3$,
and $\rho_{\rm R} \propto (1+z)^4$.  For vacuum energy, as noted above $p_{\rm
VAC}=-\rho_{\rm VAC} = -\Lambda / 8\pi G=$ constant, i.e., $w_{\rm VAC}=-1$.
For other models of dark energy, $w$ can differ from $-1$ and vary in time.
[Hereafter, $w$ without a subscript refers to dark energy.]

The present energy density of a flat Universe ($k=0$), $\rho_{\rm crit} \equiv
3H_0^2/8\pi G = 1.88\times 10^{-29} h^2 {\rm gm ~cm}^{-3} = 8.10 \times 10^{-47} h^2$ GeV$^4$, 
is known as the critical density; it provides a convenient
means of normalizing cosmic energy densities, where $\Omega_i = \rho_i(t_0) /
\rho_{\rm crit}$. 
For a positively curved Universe, $\Omega_{0} \equiv \rho(t_0) /\rho_{\rm crit} > 1$ and 
for a negatively curved Universe $\Omega_{0} <1$. The present value of the curvature radius, 
$R_{\rm curv}\equiv a/\sqrt{|k|}$, 
is related to $\Omega_0$
and $H_0$ by $R_{\rm curv} = H_0^{-1}/\sqrt{|\Omega_0 - 1|}$, 
and the characteristic scale $H_0^{-1} \approx 3000 h^{-1}$ Mpc 
is known as the Hubble radius.
Because of the evidence from the CMB that the Universe is nearly 
spatially flat (see Fig. \ref{fig:allen06}), we shall  
assume $k=0$ except where otherwise noted.

\begin{figure}[!ht]
\centerline{\psfig{file=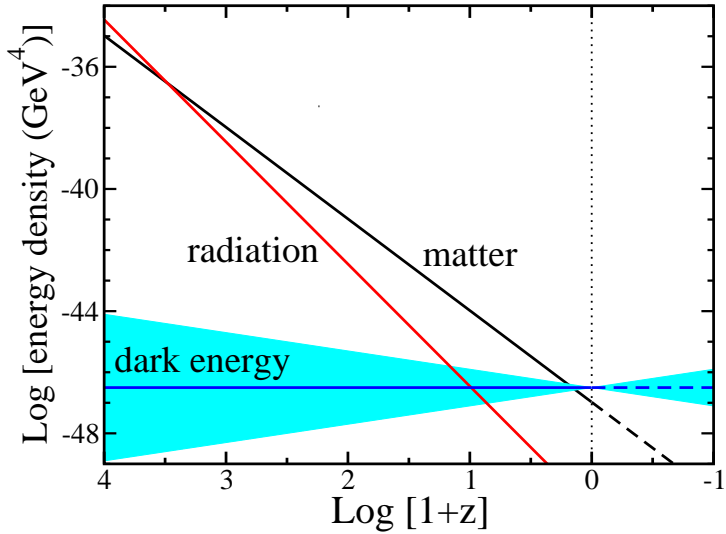,height=3.2in,angle=-90}}
\caption{Evolution of radiation, matter, and 
dark energy densities with redshift.  
For dark energy, the band represents $w=-1\pm 0.2$.  
}
\label{fig:scalings}
\end{figure}

Fig.\ \ref{fig:scalings} shows the evolution of the radiation, matter, and
dark energy densities with redshift.  
The Universe has gone through three distinct eras:
radiation-dominated, $z \gtrsim 3000$; matter-dominated, $3000
\gtrsim z \gtrsim 0.5$; and dark-energy dominated, $z \lesssim 0.5$.  The
evolution of the scale factor is controlled by the dominant energy form: $a(t)
\propto t^{2/3(1+w)}$ (for constant $w$). During the radiation-dominated era, $a(t) \propto
t^{1/2}$; during the matter-dominated era, $a(t) \propto t^{2/3}$; and for 
the dark energy-dominated era, assuming $w=-1$, asymptotically $a(t) \propto
\exp (Ht)$. For a flat Universe with matter and vacuum energy, the general 
solution, which approaches the latter two above at early and late times, is 
$a(t)=(\Omega_{\rm M}/\Omega_{\rm VAC})^{1/3} (\sinh[3\sqrt{\Omega_{\rm VAC}}H_0t/2])^{2/3}$. 

The deceleration parameter, $q(z)$, is defined as
\begin{equation}
q (z) \equiv -{\ddot a\over a H^2} = {1\over 2} \sum_i \Omega_i(z) \left[ 1+ 3w_i(z) \right]
\label{eq:q}
\end{equation}
where $\Omega_i(z)\equiv \rho_i(z)/\rho_{\rm crit}(z)$ is the fraction of critical density in component $i$ at redshift $z$.  During the matter- and radiation-dominated
eras, $w_i>0$ and gravity slows the expansion, so that $q>0$ and $\ddot a <0$.
Because of the $(\rho + 3p)$ term in the second Friedmann equation (Newtonian
cosmology would only have $\rho$), the gravity of a component that 
satisfies $p < -\rho/3$, i.e., $w<-1/3$, is repulsive and can cause
the expansion to accelerate (${\ddot a}>0$): we take this to be the defining property of
dark energy.
The successful predictions of the radiation-dominated era of 
cosmology, e.g., big bang nucleosynthesis and the formation 
of CMB anisotropies, 
provide evidence for the $(\rho + 3p)$ term, since during this epoch $\ddot a$
is about twice as large as it would be in Newtonian cosmology.

\subsection{Distances and the Hubble diagram}
 
\begin{figure}[!t]
\begin{center}
\centerline{\psfig{file=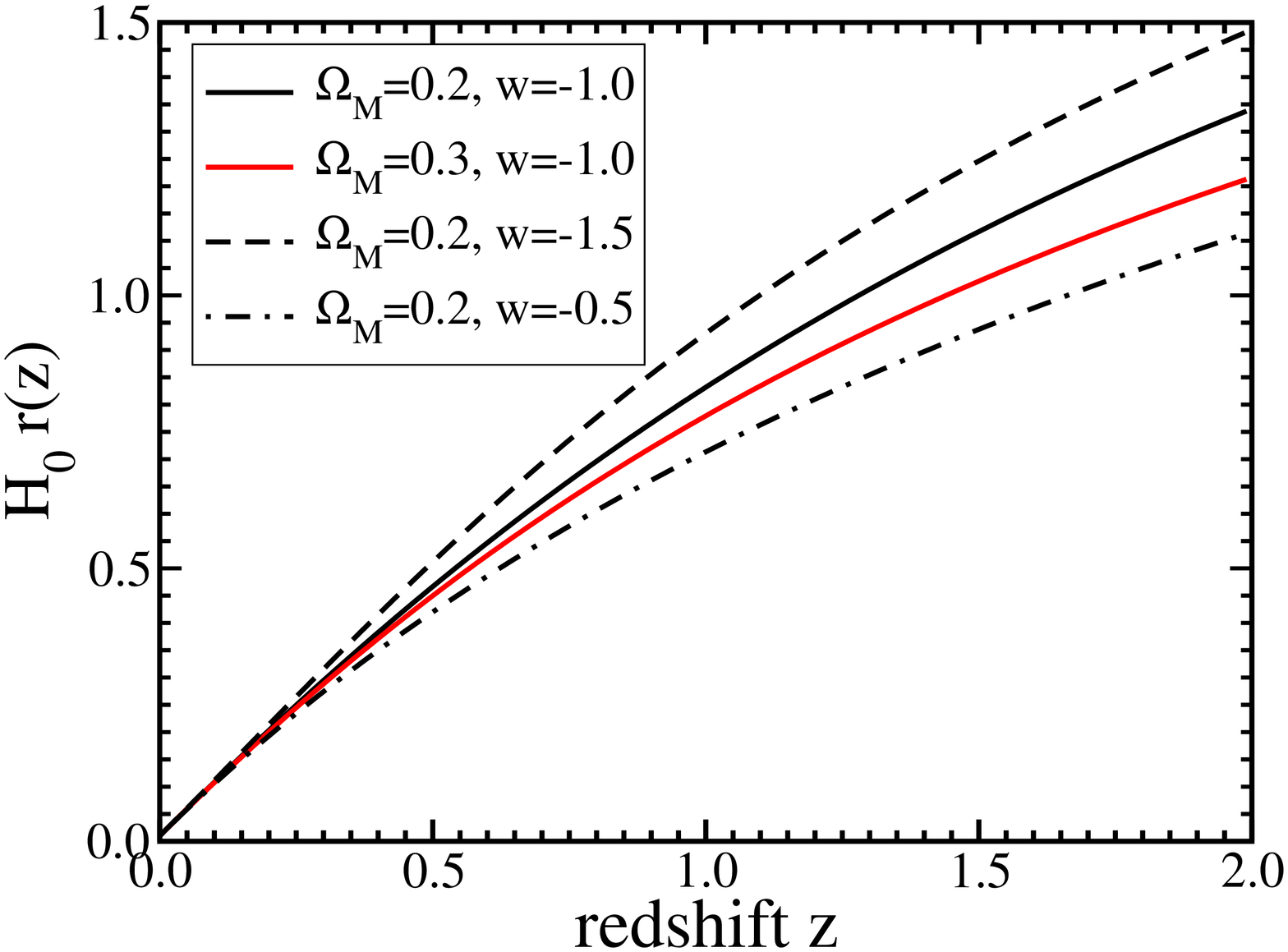,width=3in,angle=-90}
\psfig{file=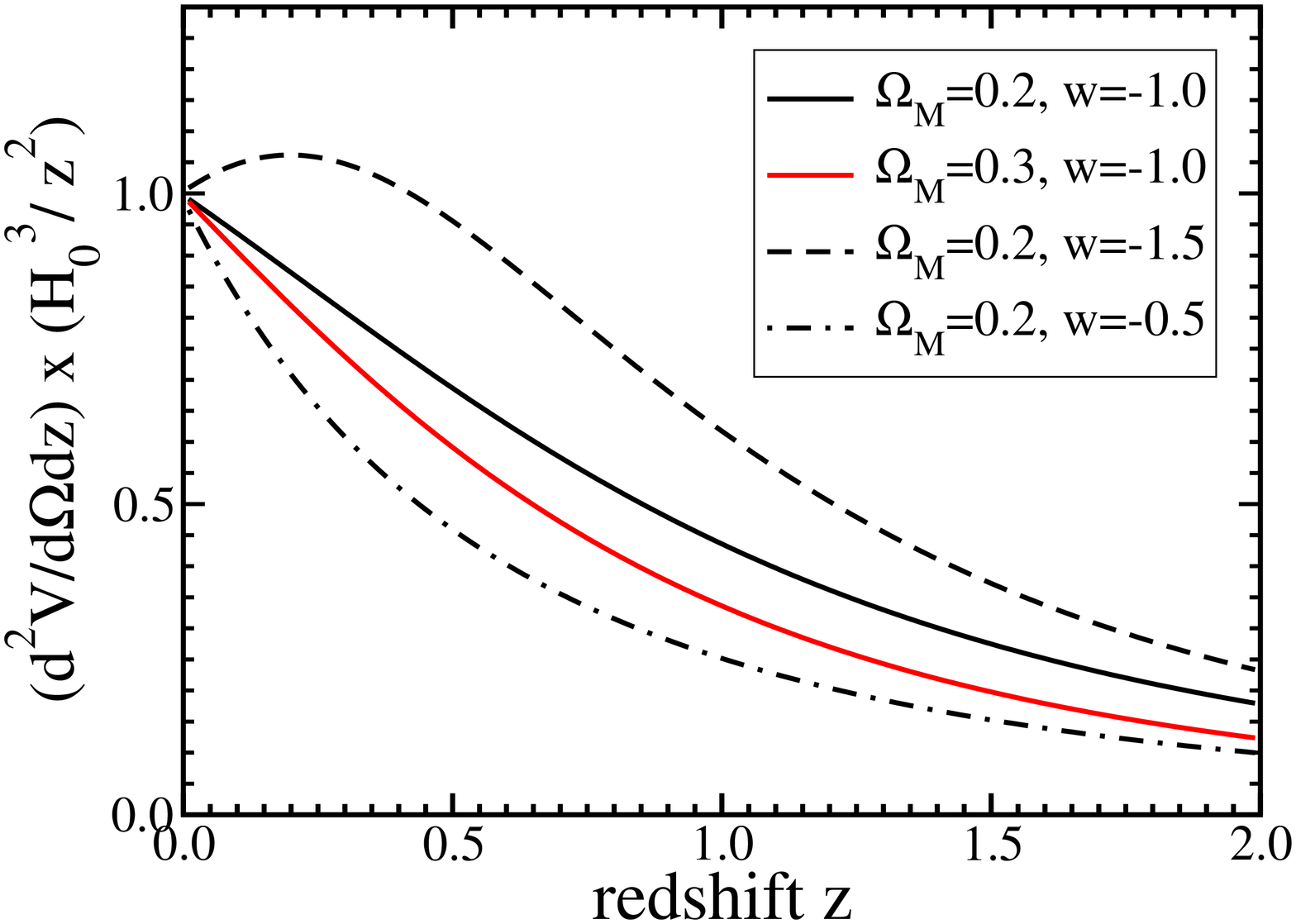,width=3in,angle=-90}
}
\caption{For a flat Universe, 
the effect of dark energy upon cosmic distance (left) and volume element 
(right) is controlled by $\Omega_{\rm M}$ and $w$.}
\label{fig:dist_vol}
\end{center}
\end{figure}

For an object of intrinsic luminosity $L$, the measured energy flux $F$
defines the luminosity distance $d_L$ to the object, i.e., the distance
inferred from the inverse square law.  The luminosity distance is related to
the cosmological model through
\begin{equation}
d_L(z) \equiv \sqrt{L \over 4 \pi F} = (1+z)r(z)~, 
\label{eq:dL}
\end{equation}
where $r(z)$ is the comoving distance to an object at redshift $z$,  
\begin{eqnarray}
r(z) & = & \int_0^z {dz'\over H(z')} = \int_{1/(1+z)}^1 {da\over a^2 H(a)}
~~~~~(k=0)~,\\
r(z) & = & |k|^{-1/2} \chi \left[ |k|^{1/2} \int_0^z dz'/H(z') \right] 
~~~~~(k\not= 0)~,
\end{eqnarray}
and where $\chi(x) = \sin (x)$ for $k>0$ and $\sinh (x)$ for $k<0$.  Specializing to
the flat model and constant $w$, 
\begin{equation}
r(z) = {1\over H_0}\int_0^z \,
{dz' \over \sqrt{\Omega_{\rm M}(1+z')^3 + (1-\Omega_{\rm M})(1+z')^{3(1+w)} + \Omega_{\rm R} (1+z')^4}}
\label{eq:rz}
\end{equation}
where $\Omega_{\rm M}$ is the present fraction of critical density in 
non-relativistic matter, and 
$\Omega_{\rm R} \simeq 0.8 \times 10^{-4}$ represents the small contribution
to the present energy density from photons and relativistic neutrinos.  
In this model, the dependence of cosmic distances upon dark energy is
controlled by the parameters $\Omega_{\rm M}$ and $w$ and 
is shown in the left panel of 
Fig.~\ref{fig:dist_vol}.

The luminosity distance is related to the distance modulus $\mu$ by
\begin{equation}
\mu(z) \equiv m-M = 5\log_{10}\left ({d_L / 10\,{\rm pc}}\right ) 
=
5 \log_{10} \left[(1+z) r(z)/{\rm pc} \right] - 5
~,
\label{eq:distmod}
\end{equation}
where $m$ is the apparent magnitude of the object (proportional to the log of
the flux) and $M$ is the the absolute magnitude (proportional to the log of the
intrinsic luminosity). 
``Standard candles,'' objects of fixed absolute magnitude $M$, and
measurements of the logarithmic energy flux $m$ constrain the cosmological model
and thereby the expansion history through this magnitude-redshift relation, 
known as the Hubble diagram. 

Expanding the scale factor around its value today, $a(t) = 1 + H_0 (t-t_0) -
q_0H_0^2 (t-t_0)^2/2 + \cdots ,$ the distance-redshift relation can be
written in its historical form
\begin{equation}
 H_0 d_L = z + {1\over 2} (1-q_0)z^2 + \cdots 
\end{equation}
The expansion rate and
deceleration rate today
appear in the first two terms in the Taylor expansion of the
relation.  This expansion, only valid for $z\ll 1$, is of
historical significance and utility; it is not useful today since objects as
distant as redshift $z \sim 2$ are being used to probe the expansion history.
However, it does illustrate the general principle: the first term on the
r.h.s. represents the linear Hubble expansion, and the deviation from a linear
relation reveals the deceleration (or acceleration).

The angular-diameter distance $d_A$, the distance inferred from the angular
size $\delta \theta$ of a distant object of fixed diameter $D$, is defined by
$d_A \equiv D/\delta \theta = r(z)/(1+z)=d_L/(1+z)^2$.  The use of ``standard
rulers'' (objects of fixed intrinsic size) provides another means of probing
the expansion history, again through $r(z)$.

The cosmological time, or time back to the Big Bang, is given by
\begin{equation}
t(z) = \int_0^{t(z)} dt'  =  \int_z^\infty {dz' \over (1+z')H(z')}~.
\label{eq:cosmic_time}
\end{equation}
While the present age in principle depends
upon the expansion rate at very early times, the rapid rise of $H(z)$ with $z$
--- a factor of 30,000 between today and the epoch of last scattering, when 
photons and baryons decoupled, at $z_{LS}\simeq 1100$, $t(z_{LS}) \simeq 380,000$ years 
--- makes this point moot.

Finally, the comoving volume element per unit solid angle $d\Omega$ is given 
by
\begin{equation}
{d^2V\over dzd\Omega} = r^2{dr\over dz}{1\over \sqrt{1-kr^2}} = {r^2(z)\over H(z)}\, .
\label{eq:volume}
\end{equation}
For a set of objects of known comoving density $n(z)$, the 
comoving volume element can be used to infer $r^2(z)/H(z)$ from 
the number counts per unit redshift and solid angle,  
$d^2N/dzd\Omega = n(z) d^2V/dzd\Omega$.  The dependence of the comoving volume
element upon $\Omega_{\rm M}$ and $w$ is shown in 
the right panel of Fig. \ref{fig:dist_vol}.

\subsection{Growth of structure and $\Lambda$CDM}
\label{sec:growth}

\begin{figure}[!ht]
\centerline{\psfig{file=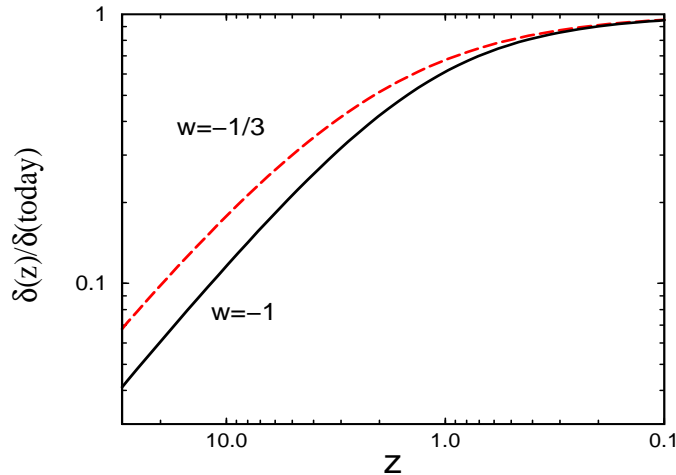,width=3.5in,height=2.6in}}
\caption{Growth of linear density perturbations in a flat universe 
with dark energy. 
Note that the growth of perturbations ceases when dark energy 
begins to dominate, $1+z =(\Omega_{\rm M}/\ode)^{1/3w}$. 
}
\label{fig:growth}
\end{figure}

A striking success of the consensus cosmology is its ability to 
account for the observed structure in the
Universe, provided that the dark matter is composed of 
slowly moving particles, known as
cold dark matter (CDM), and that the initial power 
spectrum of density perturbations
is nearly scale-invariant, $P(k) \sim k^{n_S}$ with spectral 
index $n_S \simeq 1$, as
predicted by inflation \citep{Springel06}.  Dark energy affects the
development of structure by its influence on the expansion rate of the
Universe when density perturbations are growing.  This fact and the 
quantity and quality of large-scale structure data make structure 
formation a sensitive probe of dark energy.

In GR the growth of small-amplitude, matter-density perturbations
on length scales much smaller than the Hubble radius is governed by
\begin{equation}
\ddot\delta_k + 2H\dot\delta_k -4\pi G \rho_{\rm M} \delta_k = 0\, ,
 \label{eq:growth}
\end{equation}
where the perturbations 
$\delta({\bf x},t) \equiv \delta \rho_{\rm M}({\bf x},t)/{\bar \rho_{\rm M}}(t)$ 
have been decomposed into their Fourier modes of
wavenumber $k$, and matter is assumed to be pressureless (always true for
the CDM portion and valid for the baryons on mass scales larger than
$10^5\,M_\odot$ after photon-baryon decoupling).  Dark energy affects the
growth through the ``Hubble damping'' term, $2H\dot\delta_k$. 

The solution to Eq.~(\ref{eq:growth}) is simple to describe during the three
epochs of expansion discussed earlier: $\delta_k (t)$ grows as $a(t)$ during
the matter-dominated epoch and is approximately constant during the
radiation-dominated and dark energy-dominated epochs.  The key feature here is
the fact that once accelerated expansion begins, the growth of linear
perturbations effectively ends, since the Hubble damping time becomes
shorter than the timescale for perturbation growth.  

The impact of the dark energy 
equation-of-state parameter $w$ on the growth of 
structure is more subtle and is
illustrated in Fig.~\ref{fig:growth}. For larger $w$ and fixed 
dark energy density $\Omega_{\rm DE}$,
dark energy comes to dominate earlier, causing the growth of linear
perturbations to end earlier; this means the growth factor since decoupling is
smaller and that to achieve the same amplitude by today, the perturbation must
begin with larger amplitude and is larger at all redshifts until today. The
same is true for larger $\Omega_{\rm DE}$ and fixed $w$. Finally, 
if dark energy is dynamical (not vacuum energy), then in 
principle it can be inhomogeneous, an effect ignored above. In practice, 
it is expected to be nearly uniform over scales smaller than the 
present Hubble radius, in sharp contrast to dark matter, which can 
clump on small scales.

\section{FROM EINSTEIN TO ACCELERATED EXPANSION}
\label{history}
Although the discovery of cosmic acceleration is often portrayed as a major surprise and a radical contravention of the conventional wisdom, it was anticipated by a number of developments in cosmology in the preceding decade. Moreover, this is not the first time that the cosmological constant has been proposed. Indeed, the cosmological constant was explored from the very beginnings of General Relativity and has been periodically invoked and subsequently cast aside several times since. Here we recount some of this complex 90-year history.

\subsection{Greatest blunder?}

Einstein introduced the cosmological constant in his field equations in order to obtain a static and finite cosmological solution ``as required by the fact of the small velocities of the stars'' and to be consistent with Mach's principle \citep{einstein17}. In Einstein's solution, space is positively curved, $R_{\rm curv}= 1/\sqrt{4\pi G \rho_{\rm M}}$, and the ``repulsive gravity'' of $\Lambda$ is balanced against the attractive gravity of matter, $\rho_\Lambda = \rho_{\rm M}/2$.  In the 1920's, Friedmann and Lema\^{\i}tre independently showed that cosmological 
solutions with matter and $\Lambda$ generally involved expansion or contraction, and Lema\^{\i}tre as well as Eddington showed that Einstein's static solution was unstable to expansion or contraction. In 1917, de Sitter 
explored a solution in which $\rho_{\rm M}$ is negligible compared 
to $\rho_\Lambda$ \citep{Sitter}. 
There was some early confusion about the interpretation 
of this model, 
but in the early 1920's, Weyl, Eddington, 
and others showed that the apparent recession 
velocity (the redshift) at small separation would be proportional 
to the distance, $v=\sqrt{\Lambda/3}~d$. 

With Hubble's discovery of the expansion of the Universe in 1929, 
Einstein's primary justification for introducing the cosmological 
constant was lost, and he advocated abandoning it. 
Gamow later wrote that Einstein called this ``his 
greatest blunder,'' since he could have predicted the expanding Universe. 
Yet the description above makes it clear that the history was 
more complicated, and one could argue that in fact Friedmann and Lema\^{\i}tre (or de Sitter)
{\it had} ``predicted'' the expanding Universe, $\Lambda$ or no. Indeed, 
Hubble noted that his linear relation between redshift 
and distance was consistent with the prediction of the de Sitter model 
\citep{Hubble29}.  Moreover, Eddington 
recognized that Hubble's value for the expansion rate, 
$H_0 \simeq 570\,$km/s/Mpc, implied a time back to the big bang of less than 2\,Gyr, 
uncomfortably short compared to some age estimates of Earth and the galaxy.  
By adjusting the cosmological constant to be slightly larger than the Einstein
value, $\rho_\Lambda = (1+\epsilon )\rho_{\rm M}/2$, a nearly static beginning of
arbitrary duration could be obtained, a solution known as the 
Eddington-Lema\^{\i}tre model. While Eddington remained focused on $\Lambda$, 
trying to find a place for it in his ``unified'' and ``fundamental'' theories, 
$\Lambda$ was no longer the focus of most cosmologists.

\subsection{Steady state and after}

Motivated by the aesthetic beauty of an unchanging Universe, 
\citet{bondi} and \citet{hoyle} put forth the steady-state cosmology, 
a revival of the de Sitter model with a new twist.  
In the steady-state model, the dilution of matter due to expansion is
counteracted by postulating the continuous creation of matter (about 1
hydrogen atom/m$^3$/Gyr).  However, the model's firm prediction of an unevolving
Universe made it
easily falsifiable, and the redshift distribution of radio galaxies, the absence
of quasars nearby, and the discovery of the cosmic microwave background
radiation did so in the early 1960s.

$\Lambda$ was briefly resurrected again in the late 1960s by
\citet{Petrosian}, who used the Eddington-Lema\^{\i}tre model
to explain the preponderance of quasars at redshifts around $z \sim 2$. As it
turns out, this is a real observational effect, but it can be attributed to
evolution: quasar activity peaks around this redshift.  In 1975, evidence for
a cosmological constant from the Hubble diagram of brightest-cluster
elliptical galaxies was presented \citep{Gunn_Tinsley}, though it was realized
\citep{Tinsley_Gunn} that uncertainties in 
galaxy luminosity evolution make their use as standard candles problematic.

While cosmologists periodically hauled the cosmological constant  
out of the closet as needed and then stuffed it back in, in the 1960s 
physicists 
began to understand that $\Lambda$ cannot be treated in such cavalier fashion. 
With the rise of the standard big-bang cosmology came the awareness that 
the cosmological constant could be a big problem \citep{zeldovich68}.  
It was realized that the energy density 
of the quantum vacuum should result in a cosmological constant of enormous size (see \S \ref{sec:vacuum}).  However, because of the success of the hot big-bang model, the 
lack of compelling ideas to solve the cosmological constant problem, 
and the dynamical unimportance of $\Lambda$ at the early epochs when the 
hot big-bang model was best tested by big-bang nucleosynthesis and the CMB, 
the problem was largely ignored in cosmological discourse.

\subsection{Enter inflation}
In the early 1980s the inflationary universe scenario \citep{guth80}, with its predictions of a
spatially flat Universe ($\Omega=1$) and almost-scale-invariant density perturbations, changed the cosmological landscape and helped set the stage for the discovery of cosmic acceleration.
When inflation was first introduced, the evidence
for dark matter was still accruing, and estimates of the total matter density, then about $\Omega_{\rm M}\sim 0.1$, were sufficiently uncertain that an Einstein-de Sitter model (i.e., $\Omega_{\rm M} = 1$) was not ruled out. The evidence for a low value of $\Omega_{\rm M}$ was, however, sufficiently worrisome that the need for a smooth component, such as vacuum energy, to make up the difference for a flat Universe was suggested \citep{peebles84,turner84}.  Later, the model for large-scale structure formation with a cosmological constant and cold dark matter ($\Lambda$CDM) and the spectrum of density perturbations predicted by inflation was found to provide a better fit (than $\Omega_{\rm M}=1$) to the growing observations of large-scale structure \citep{Turner91,efstathiou90}.  The 1992 COBE discovery of CMB anisotropy provided the normalization of the spectrum of density perturbations and drove a spike into the heart of the $\Omega_M =1$ CDM model.

Another important thread involved age consistency. While estimates of the 
Hubble parameter had ranged between 50 and 100 km/s/Mpc since the 1970s, 
by the mid-1990s they were settling out in the middle of that range. 
Estimates of old globular cluster ages had similar swings, but had
settled at $t_0 \simeq 13-15$ Gyr. The resulting expansion age, 
$H_0 t_0 = (H_0/70~{\rm km/s/Mpc})(t_0/14~{\rm Gyr})$ was uncomfortably 
high compared to that for the Einstein-de Sitter model, 
for which $H_0 t_0=2/3$. The cosmological constant offered a ready solution, as the age of a flat Universe with $\Lambda$ rises with $\Omega_\Lambda$, 
\begin{equation}
H_0 t_0 = {1\over 3\Omega_\Lambda^{1/2}}
\ln \left[ {1+{\Omega_\Lambda}^{1/2}\over 1-{\Omega_\Lambda}^{1/2}} \right] 
= {2\over 3} \left[ 1 + {\Omega_\Lambda}^2 /3 + {\Omega_\Lambda}^4 /5 + 
\cdots \right] ~, 
\label{eq:H0t0}
\end{equation}
reaching $H_0 t_0 \simeq 1$ for $\Omega_\Lambda=0.75$. 

By 1995 the cosmological constant was back out 
of the cosmologists' closet in full glory \citep{Krauss_Turner,JPO_PJS,Frieman_PNGB}:  it solved the age problem, was consistent with growing evidence that $\Omega_M$ was around 0.3, and fit the growing body of observations of large-scale structure.  Its only serious competitors were ``open inflation,'' which had a small group of adherents, and hot + cold dark matter, with a low value 
for the Hubble parameter ($\sim 50$\,km/s/Mpc) and neutrinos accounting for 10\% to 15\% of the dark matter \citep[see, e.g., contributions in][]{CriticalDialogues}.
During this period, 
there were two results that conflicted with $\Lambda$CDM: analysis of 
the statistics of lensed quasars 
\citep{Kochanek_96} and of the first 7 high-redshift supernovae 
of the Supernova Cosmology Project 
\citep{Perlmutter_97} respectively 
indicated that $\Omega_\Lambda < 0.66$ and $\Omega_\Lambda < 0.51$ at 95\% confidence, for 
a flat Universe.  The  
discovery of accelerated expansion in 1998 saved inflation by providing 
evidence for large $\Omega_\Lambda$ and  
was thus welcome news for cosmology.

\begin{figure}[!h]
\centerline{\psfig{file=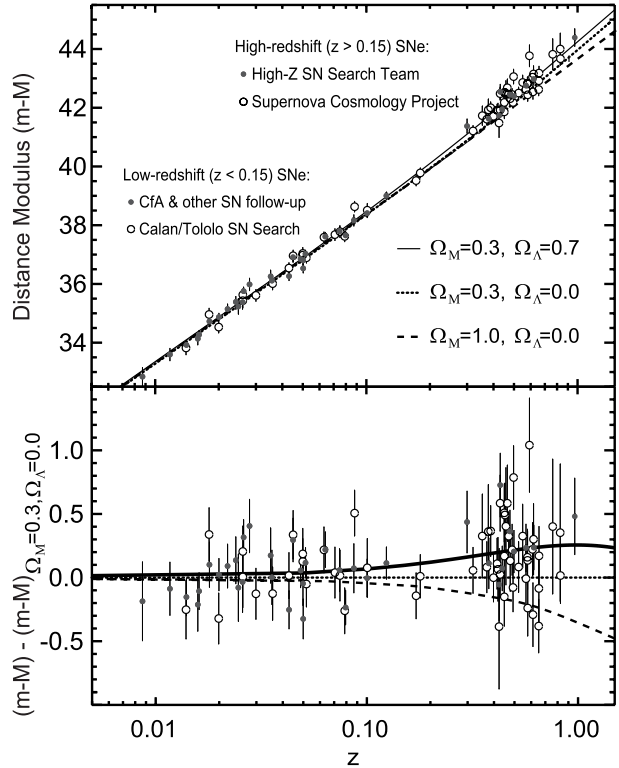,width=3.2in}}
\caption{Discovery data: Hubble diagram of SNe Ia measured by the Supernova
Cosmology Project and the High-z Supernova Team. Bottom panel shows 
residuals in distance modulus 
relative to an open universe with $\Omega_{0} = \Omega_{\rm M}=0.3$. Figure 
adapted from \citet{Riess_2000,Perlmutter_Schmidt}, based on 
\citet{riess98,perlmutter99}.  
}
\label{fig:hub_diag}
\end{figure}

\subsection{Discovery}

Two breakthroughs enabled the discovery of cosmic acceleration.
The first was the demonstration that type Ia supernovae (SNe Ia)
are standardizable candles
\citep{Phillips_93}.  The second was the deployment of large 
mosaic CCD cameras on 4-meter class telescopes, enabling 
the systematic search of large areas 
of sky, containing thousands of galaxies, for these rare events. 
By comparing deep, wide images taken weeks apart, 
the discovery of
SNe at redshifts $z \sim 0.5$ could be ``scheduled'' on a statistical basis.

Two teams, the Supernova Cosmology Project and the High-z SN Search,
working independently in the mid- to late-1990s took advantage of these
breakthroughs to measure the SN Hubble diagram to much larger distances 
than was previously possible.
Both teams found that distant SNe are $\sim 0.25$ mag dimmer than they
would be in a decelerating Universe, indicating that the expansion has 
been speeding up for the past 5 Gyr \citep{riess98, perlmutter99}; see
Fig.~\ref{fig:hub_diag}. When analyzed assuming 
a Universe with matter and cosmological
constant, their results provided evidence for $\Omega_{\Lambda}>0$ at greater
than 99\% confidence (see Fig.~\ref{fig:allen06} for the current 
constraints).

\section{CURRENT STATUS}
\label{currentstatus}
Since the supernova discoveries were announced in 1998, the evidence for an
accelerating Universe has become substantially stronger and more broadly
based.  Subsequent supernova observations have reinforced the original
results, and new evidence has accrued from other observational probes. In this
section, we review these developments and discuss the current status of the
evidence for cosmic acceleration and what we know about dark energy. In \S
\ref{probes}, we address the probes of cosmic acceleration in more detail, and
we discuss future experiments in \S \ref{projects}.

\subsection{Cosmic microwave background and large-scale structure}
An early and important confirmation of accelerated expansion was the
independent evidence for dark energy from measurements of CMB anisotropy 
\citep{Boom,DASI} and of 
large-scale structure (LSS).  
The CMB constrains the amplitude of the primordial
fluctuations that give rise to the observed structure as well as the distance
to the last-scattering surface, $r(z\simeq 1100)$.  In order to allow
sufficient growth of the primordial perturbations and not disrupt the
formation of large-scale structure, dark energy must come to dominate the
Universe only very recently (see \S \ref{sec:growth}), implying that its energy
density must evolve with redshift more slowly than matter.  This occurs if it
has negative pressure, $w<0$, cf. Eq.~(\ref{eq:rhow}).  Likewise, the presence of a component 
with large negative pressure that accounts for three-quarters
of the critical density affects the distance to the last-scattering surface.

\subsubsection{CMB}
\begin{figure}[!t]
\begin{center}
\centerline{\psfig{file=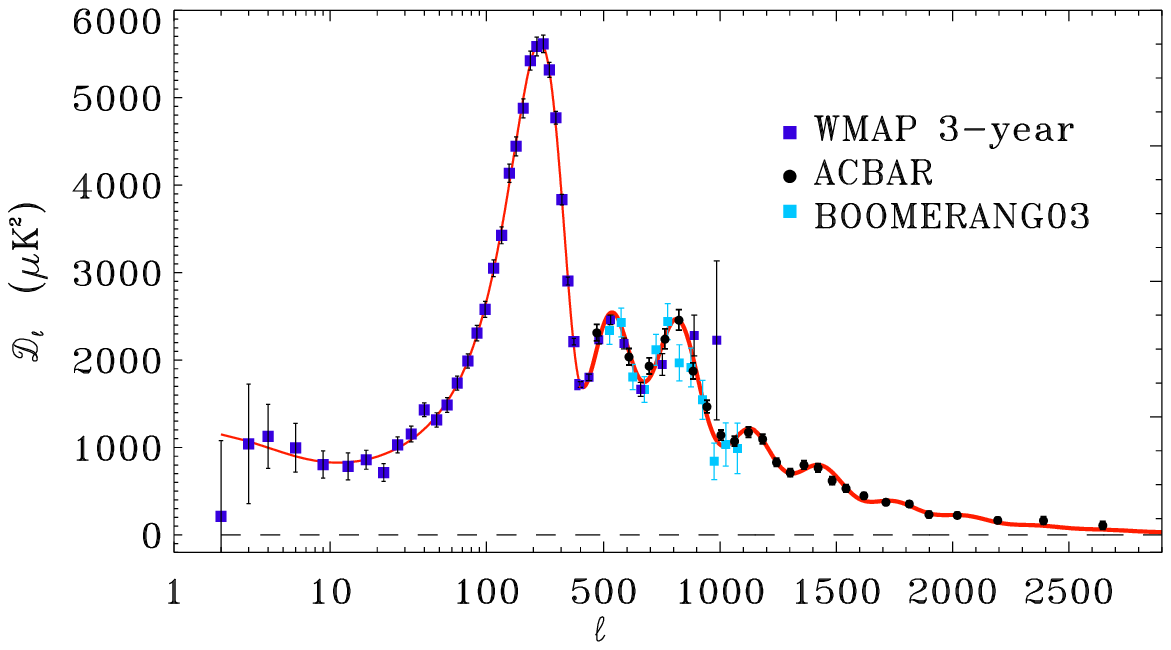,width=3.in,height=2.2in}
\hspace*{0.1in}
\psfig{file=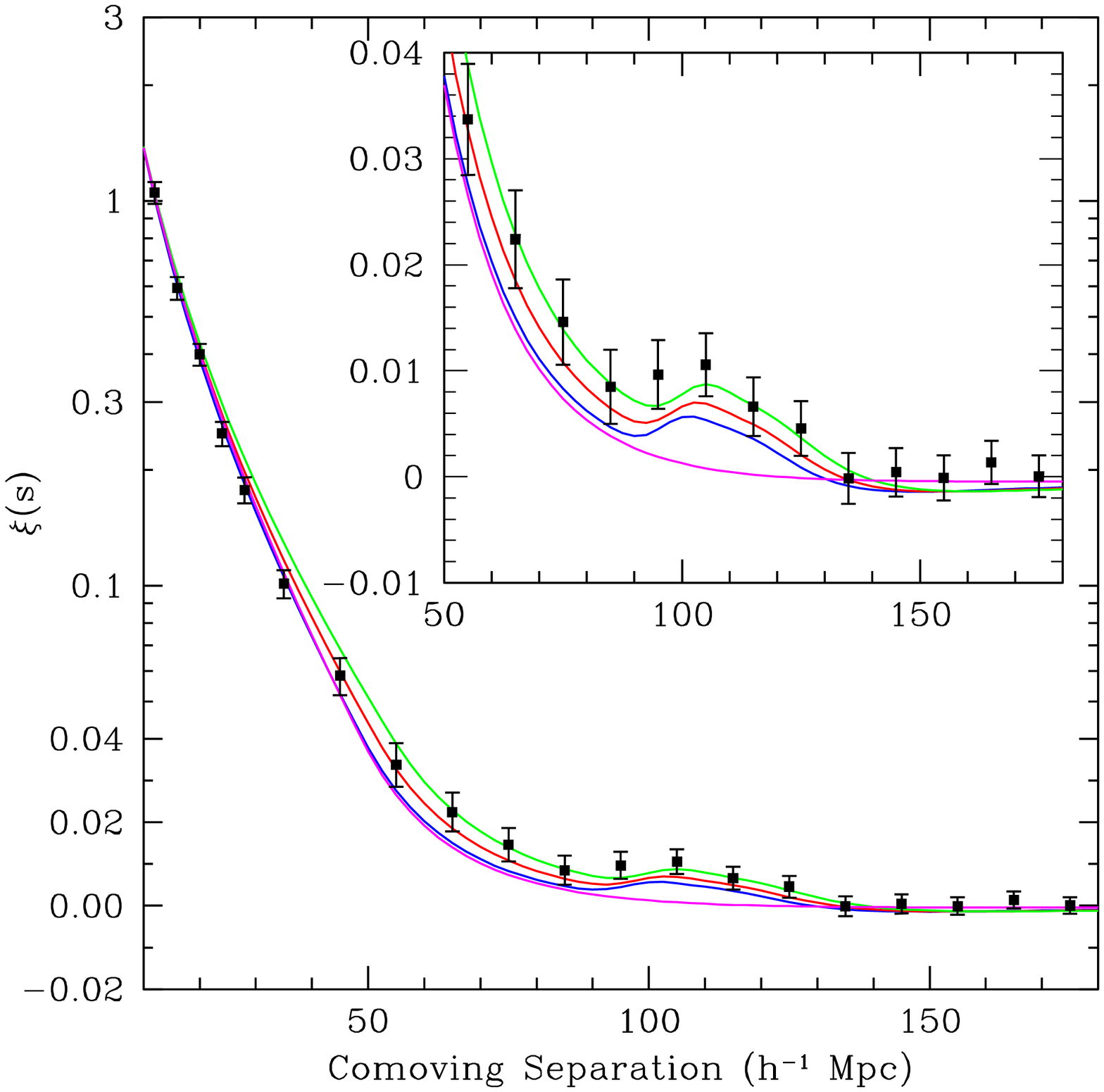,width=2.3in}}
\caption{{\it Left panel:} Angular power spectrum measurements of the 
CMB temperature fluctuations from WMAP, Boomerang, and ACBAR. 
Red curve shows the best-fit $\Lambda$CDM model.  From \citet{Reichardt08}.
{\it Right panel:} Detection of the
baryon acoustic peak in the clustering of luminous red galaxies in the SDSS
\citep{SDSS_BAO}.  Shown is the two-point galaxy correlation function in
redshift space; inset shows an expanded view with a linear vertical axis.
Curves correspond to $\Lambda$CDM predictions for $\Omega_{\rm M} h^2=0.12$ (green),
$0.13$ (red), and $0.14$ (blue). Magenta curve shows a $\Lambda$CDM model
without BAO. }
\end{center}
\label{fig:CMB}
\end{figure}

Anisotropies of the cosmic microwave background provide a record of the
Universe at a simpler time, before structure had developed and when photons
were decoupling from baryons, about 380,000 years after the Big Bang
\citep{Hu_Dodelson}.  The angular power spectrum of CMB temperature
anisotropies, measured most recently by WMAP \citep{WMAP_3} and by
ground-based experiments that probe to smaller angular scales, is dominated by
acoustic peaks that arise from gravity-driven sound waves in the photon-baryon
fluid (see Fig.\ 5a).  The positions and amplitudes of the acoustic peaks
encode a wealth of cosmological information.  They indicate
that the Universe is nearly spatially flat to within a few percent. 
In combination with LSS or with independent $H_0$ measurement, 
the CMB measurements indicate that 
matter contributes only about a quarter of the critical density.  A component
of missing energy that is smoothly distributed 
is
needed to square these observations -- and is fully consistent with the dark
energy needed to explain accelerated expansion.

\subsubsection{Large-scale Structure}

Baryon acoustic oscillations (BAO), so prominent in the CMB anisotropy, leave
a subtler characteristic signature in the clustering of galaxies, a bump in
the two-point correlation function at a scale $\sim 100$ Mpc that can be
measured today and in the future can provide a powerful probe of dark energy
(see \S \ref{sec:BAO}).  Measurement of the BAO signature in the correlation
function of SDSS luminous red galaxies (see Fig.\ 5b) constrains the distance to
redshift $z=0.35$ to a precision of 5\% \citep{SDSS_BAO}.  This measurement
serves as a significant complement to other probes, as shown in
Fig.~\ref{fig:allen06}.

The presence of dark energy affects the large-angle anisotropy of the CMB (the
low-$\ell$ multipoles) through the integrated Sachs-Wolfe (ISW) effect.  The
ISW arises due to the differential redshifts of photons as they pass through
time-changing gravitational potential wells, and it leads to a small correlation
between the low-redshift matter distribution and the CMB anisotropy.  This
effect has been observed in the cross-correlation of the CMB with
galaxy and radio source catalogs
\citep{Boughn_Crittenden,Fosalba_Gaztanaga,Afshordi_Strauss,Scranton_ISW}. This 
signal indicates that the Universe is not described by the Einstein-de Sitter
model ($\Omega_{\rm M}=1$), a reassuring cross-check.

Weak gravitational lensing \citep{Schneider_review,Munshi_review}, the small,
correlated distortions of galaxy shapes due to gravitational lensing by
intervening large-scale structure, is a powerful technique for mapping dark
matter and its clustering. Detection of this cosmic shear signal was first
announced by four groups in 2000 \citep{bacon00,kaiser00,vanW00,Witt00}.
Recent lensing surveys covering areas of order 100 square degrees have shed
light on dark energy by pinning down the combination $\sigma_8
(\Omega_{\rm M}/0.25)^{0.6}\approx 0.85\pm 0.07$, where $\sigma_8$ is the rms
amplitude of mass fluctuations on the $8~h^{-1}$ Mpc scale
\citep{Jarvis_CTIO,Hoekstra,Massey}.  Since other measurements peg
$\sigma_8$ at $\simeq 0.8$, 
this implies that $\Omega_{\rm M} \simeq 0.25$, consistent with
a flat Universe dominated by dark energy. In the future, weak lensing has the
potential to be the most powerful probe of dark energy
\citep{Huterer_thesis,Hu02}, and this is discussed in \S \ref{probes} and \S
\ref{projects}.

\begin{figure}[!ht]
\centerline{\psfig{file=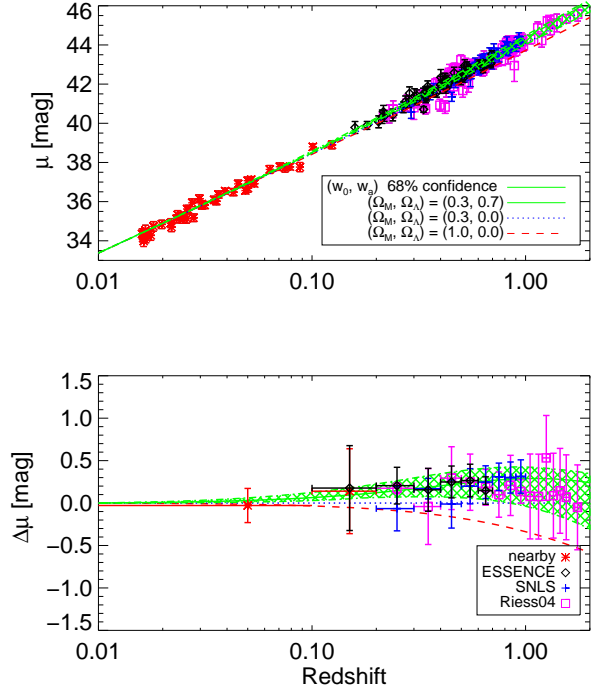,width=3.4in}}
\caption{SN Ia results: ESSENCE (diamonds), 
SNLS (crosses), low-redshift SNe (*), and the compilation of \citet{Riess_04}
which includes many of the other published SN distances plus those from HST
(squares). {\it Upper:} distance modulus vs.  redshift measurements shown
with three cosmological models: $\Omega_{\rm M}=0.3$, $\Omega_\Lambda =0$ 
(dotted); $\Omega_{\rm M} =1$, $\Omega_\Lambda=0$ (dashed); and 
the 68\% CL allowed
region in the $w_0$-$w_a$ plane, assuming spatial flatness and a prior of
$\Omega_{\rm M}=0.27\pm 0.03$ (hatched). {\it Lower:} binned distance modulus
residuals from the $\Omega_{\rm M}=0.3$, $\Omega_\Lambda=0$ model.  Adapted from
\citet{WoodVasey_07}.  }
\label{fig:WV07}
\end{figure}

\subsection{Recent supernova results}

A number of
concerns were raised about the robustness of the first SN evidence 
for acceleration, e.g., it was suggested that distant SNe could 
appear fainter due to extinction by hypothetical 
grey dust rather than acceleration 
\citep{Drell00,Aguirre99}.  Over the intervening decade, the supernova
evidence for acceleration has been strengthened by results from a series of SN
surveys. Observations with the Hubble Space Telescope (HST) have provided
high-quality light curves \citep{Knop03} and have extended SN measurements to
redshift $z\simeq 1.8$, providing evidence for the expected earlier epoch of
deceleration and disfavoring dust extinction as an alternative 
explanation to acceleration \citep{Riess_01,Riess_04,Riess_06}. 
Two large ground-based surveys, the
Supernova Legacy Survey (SNLS) \citep{SNLS} 
and the ESSENCE survey 
\citep{Miknaitis_07}, have been using 4-meter telescopes to measure 
light curves for several hundred SNe Ia over the redshift range $z \sim
0.3-0.9$, with large programs of spectroscopic follow-up on 6- to 10-m
telescopes. Fig. \ref{fig:WV07} shows a compilation of SN distance
measurements from these and other surveys. The quality and quantity 
of the distant SN data are now vastly superior to what was available in 
1998, and the evidence for acceleration is correspondingly more 
secure (see Fig. \ref{fig:allen06}).

\subsection{X-ray clusters}

Measurements of the ratio of X-ray emitting gas to total mass in
galaxy clusters, $f_{\rm gas}$, also indicate the presence of dark energy.
Since galaxy clusters are the largest collapsed objects in the universe, the
gas fraction in them is presumed to be constant and nearly equal to the baryon
fraction in the Universe, $f_{\rm gas} \approx \Omega_{\rm B}/\Omega_{\rm M}$ (most of the
baryons in clusters reside in the gas).  The value of $f_{\rm gas}$ inferred
from observations depends on the observed X-ray flux and temperature as well
as the distance to the cluster.  Only the ``correct cosmology'' will produce
distances which make the apparent $f_{\rm gas}$ constant in redshift. Using
data from the Chandra X-ray Observatory, \citet{Allen_04,Allen_07} determined
$\Omega_\Lambda$ to a 68\% precision of about $\pm 0.2$, obtaining a value 
consistent with the SN data.

\subsection{Age of the Universe}

\begin{figure}[!t]
\centerline{\psfig{file=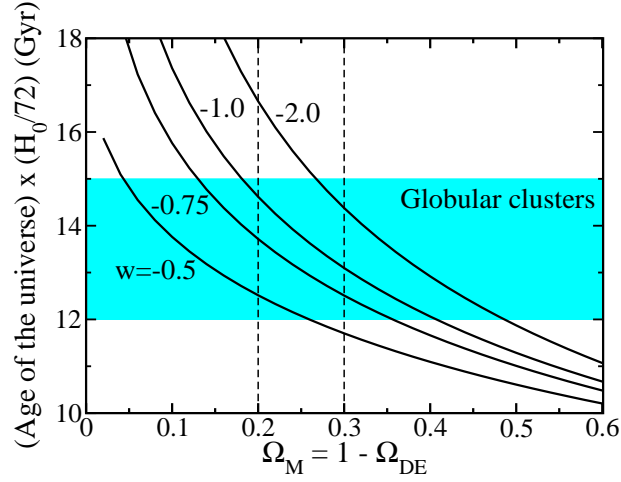,width=3.4in,angle=-90}}
\caption{Expansion age of a flat universe vs. $\Omega_{\rm M}$ for different values of $w$. 
Shown in blue are age constraints from globular clusters 
\citep{Krauss_Chaboyer}, 
and vertical dashed lines indicate 
the favored range for $\Omega_{\rm M}$. Age
consistency obtains for $-2\lesssim w\lesssim -0.5$.}
\label{fig:age}
\end{figure}

Finally, because the expansion age of the Universe depends upon the expansion
history, the comparison of this age with 
independent age estimates can be used to probe dark energy.  The ages of the
oldest stars in globular clusters constrain the age of the Universe: $12\,{\rm
Gyr}\lesssim t_0\lesssim 15$ Gyr \citep{Krauss_Chaboyer}.  
When combined with a weak constraint from structure formation or 
from dynamical measurements of the matter density, 
$0.2 < \Omega_{\rm M}<0.3$, a consistent age is
possible if $-2\lesssim w\lesssim -0.5$; see 
Fig.~\ref{fig:age}.  Age consistency is an important
crosscheck and provides additional evidence for the defining feature of dark
energy, large negative pressure. CMB anisotropy is
very sensitive to the expansion age; in combination with large-scale structure
measurements, for a flat Universe it yields the tight constraint $t_0=13.8 \pm
0.2$ Gyr \citep{SDSS_LRG}.

\subsection{Cosmological parameters}
\label{sec:combine}

\begin{table*}
  \begin{minipage}{5in}
\begin{center}
    \caption{Cosmological parameter constraints from \citet{SDSS_LRG}.}\vspace{0.2cm}
    \label{tab:cosmo_parameters}
    \begin{tabular}{ccc} \hline\hline
\rule[-3mm]{0mm}{8mm} Parameter & Consensus model & Fiducial model \\ \hline
$\Omega_0$       & $1.003\pm 0.010$       & $1$ (fixed)                  \\
$\Omega_{\rm DE}$ & $0.757 \pm 0.021$         & $0.757\pm 0.020$        \\
$\Omega_{\rm M}$       & $0.246 \pm 0.028$      & $0.243\pm 0.020$        \\
$\Omega_{\rm B}$       & $0.042 \pm 0.002$      & $0.042\pm 0.002$      \\
$\sigma_8$       & $0.747 \pm 0.046$      & $0.733\pm 0.048$        \\
$n_S$            & $0.952 \pm 0.017$      & $0.950\pm 0.016$        \\
$H_0$ (km/s/Mpc) & $72 \pm 5$             & $72\pm 3$             \\
$T_0$ (K)        & $2.725 \pm 0.001$      & $2.725 \pm 0.001$ \\
$t_0$ (Gyr)      & $13.9\pm 0.6$          & $13.8\pm 0.2$         \\
$w$              & $-1$ (fixed)           & $-0.94\pm 0.1$        \\
$q_0$            & $-0.64 \pm 0.03$       & $-0.57\pm 0.1$       \\
\hline\hline
    \end{tabular}
\end{center}
  \end{minipage}
\end{table*}

\citet{Sandage70} once 
described cosmology as the quest for two numbers, $H_0$ and
$q_0$, which were just beyond reach.  Today's 
cosmological model is described by anywhere from 4 to 20 parameters, and the
quantity and quality of cosmological data described above enables precise
constraints to be placed upon all of them. However, the results depend on {\it
which} set of parameters are chosen to describe the Universe as well as the
mix of data used.

For definiteness, we refer to the {\em ``consensus cosmological model''} (or
$\Lambda$CDM) as one in which $k$, $H_0$, $\Omega_{\rm B}$, $\Omega_{\rm M}$,
$\Omega_\Lambda$, $t_0$, $\sigma_8$, and $n_S$ are free parameters,
but dark energy is {\em assumed} to be a cosmological constant, $w=-1$. For
this model, 
\citet{SDSS_LRG} combined data from SDSS and WMAP to derive
the constraints shown in the second column of 
Table \ref{tab:cosmo_parameters}.

To both illustrate and gauge the sensitivity of the results to the choice of
cosmological parameters, we also consider a {\em ``fiducial dark energy
model''},  
in which
spatial flatness ($k=0$, $\Omega_0=1$) is imposed, and $w$ is assumed to be a
constant that can differ from $-1$.  For this case the cosmological parameter
constraints are given in the third column of Table \ref{tab:cosmo_parameters}.

\begin{figure}[!t]
\centerline{
\psfig{file=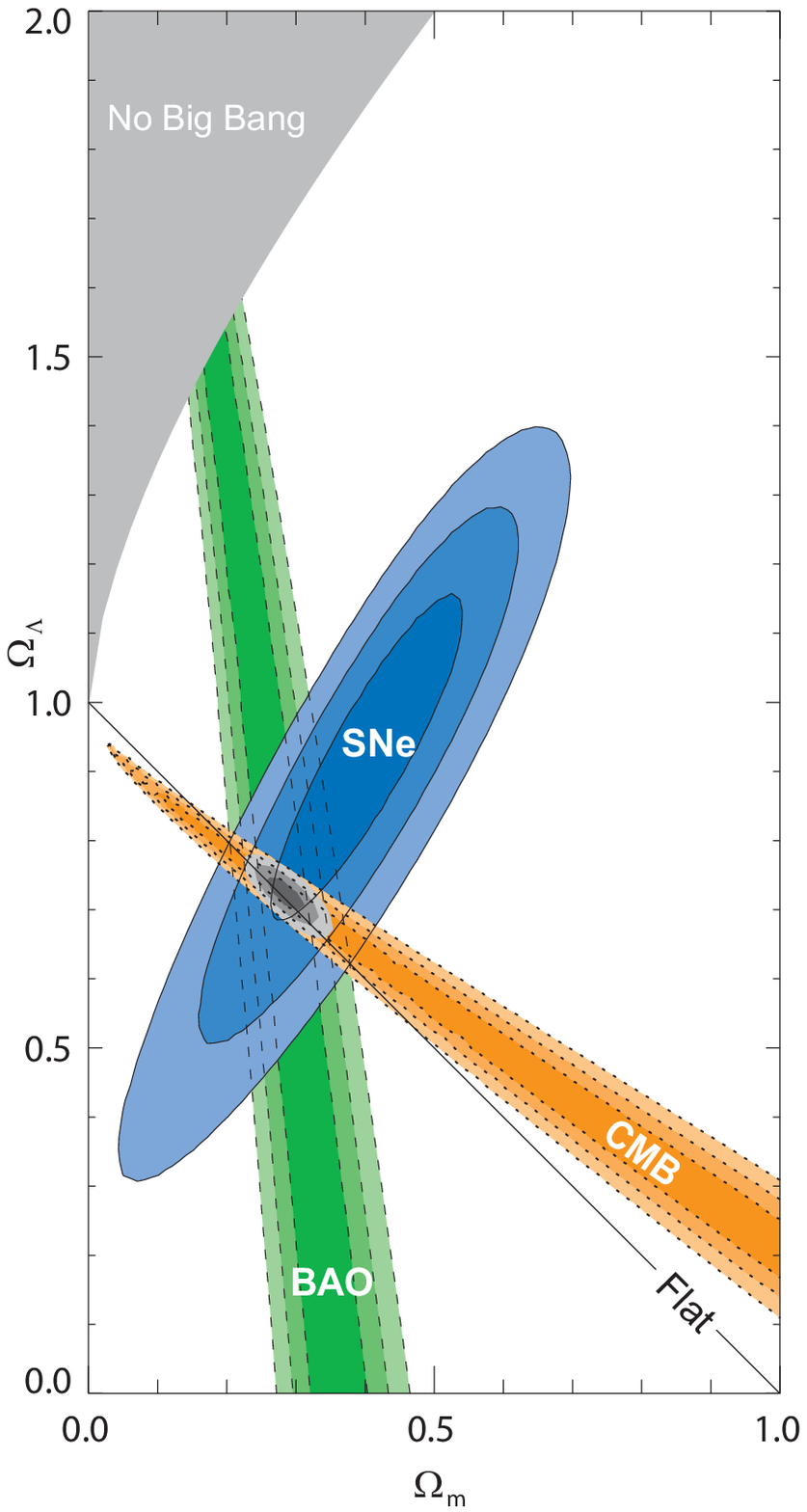,width=2in,height=3.2in}
\hspace{-1.2cm}
\vspace{-0.3cm}
\psfig{file=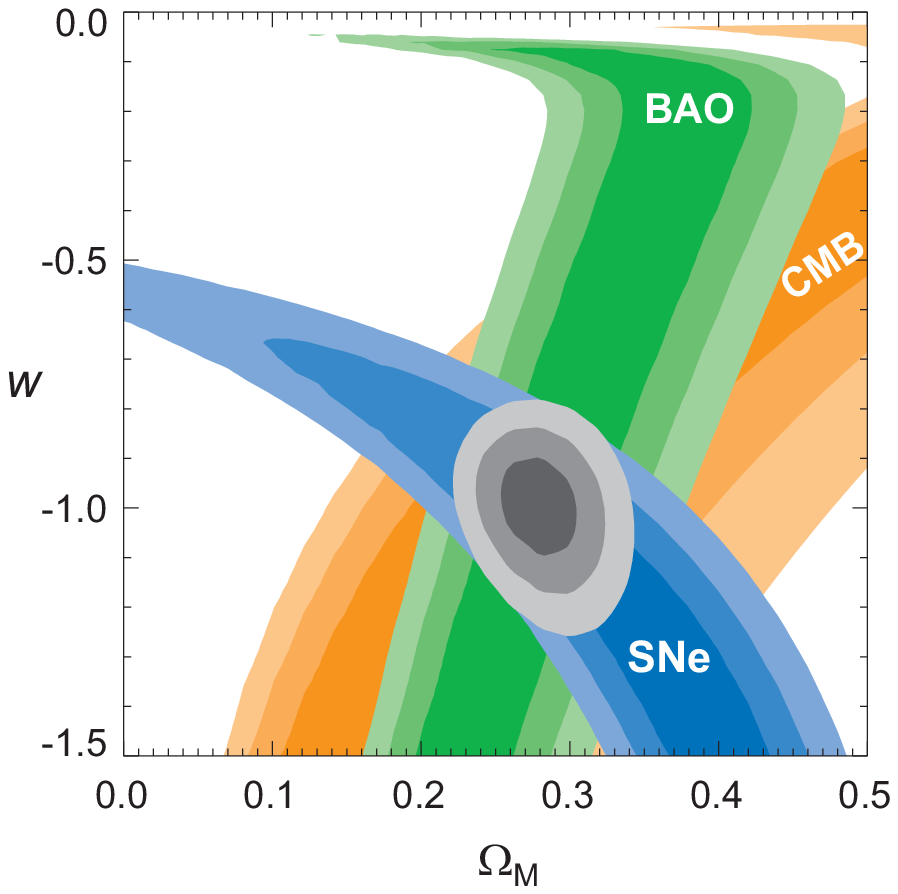,width=3.2in}
}
\caption{{\it Left panel}: Constraints upon 
$\Omega_{\rm M}$ and $\Omega_\Lambda$ in the consensus model using 
BAO, CMB, and SNe measurements.  {\it Right panel}: Constraints upon
$\Omega_{\rm M}$ and constant $w$ in the fiducial dark energy model using 
the same data sets. From \citet{Kowalski_SCP}.}
\label{fig:allen06}
\end{figure}

Although $w$ is not assumed to be $-1$ in the fiducial model, 
the data prefer a value that is consistent with this, $w = -0.94 \pm 0.1$.
Likewise, the data prefer spatial flatness in the consensus model in which
flatness is not imposed. For the other parameters, the differences are small. 
Fig.\ \ref{fig:allen06} shows how different data sets
individually and in combination constrain parameters in these two models;
although the mix of data used here differs from that in Table
\ref{tab:cosmo_parameters} (SNe are included in Fig. \ref{fig:allen06}), 
the resulting constraints are consistent.

Regarding Sandage's two numbers, Table \ref{tab:cosmo_parameters} reflects
good agreement with but a smaller uncertainty than the direct $H_0$
measurement based upon the extragalactic distance scale, $H_0 =72 \pm
8\,$km/s/Mpc \citep{Freedmanetal}. However, the parameter values in Table
\ref{tab:cosmo_parameters} are predicated on the correctness of the CDM
paradigm for structure formation.  The entries for $q_0$ in Table
\ref{tab:cosmo_parameters} are derived from the other parameters using
Eq.~(\ref{eq:q}).  Direct determinations of $q_0$ require either ultra-precise
distances to objects at low redshift or precise distances to objects at
moderate redshift.  The former is still beyond reach, while for the latter the
$H_0$/$q_0$ expansion is not valid.

If we go beyond the restrictive assumptions of these two models, allowing both
curvature and $w$ to be free parameters, then the parameter values shift
slightly and the errors increase, as expected. In this case,
combining WMAP, SDSS, 2dFGRS, and SN Ia data, \citet{WMAP_3} find $w=-1.08\pm
0.12$ and $\Omega_0= 1.026^{+0.016}_{-0.015}$, while WMAP+SDSS only bounds 
$H_0$ to the range $61-84$ km/s/Mpc at 95\% confidence \citep{SDSS_LRG}, 
comparable to the accuracy of the HST Key Project measurement \citep{Freedmanetal}.
 
\begin{figure}[!t]
\centerline{\psfig{file=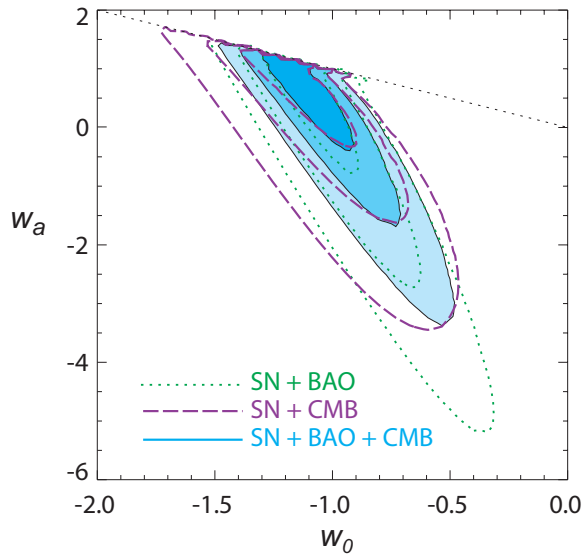,width=3.in}}
\caption{68.3\%, 95.4\%, and 99.7\% C.L. marginalized constraints on $w_0$ and $w_a$ in
a flat Universe, using data from SNe, CMB, and BAO. The diagonal line
indicates $w_0+w_a=0$.  From \citet{Kowalski_SCP}.  }
\label{fig:w0_wa}
\end{figure}

Once we drop the assumption that $w=-1$, there are no
strong theoretical 
reasons for restricting attention to constant $w$. A widely used and
simple form that accommodates evolution is $w = w_0 + (1-a)w_a$ (see \S
\ref{describingDE}).  Future surveys with greater reach than that of present
experiments will aim to constrain models in which 
$\Omega_{\rm M}, \Omega_{\rm DE},
w_0$, and $w_a$ are all free parameters (see \S \ref{projects}). We note that
the current observational constraints on such models are quite weak. 
Fig.\ \ref{fig:w0_wa} shows the marginalized constraints on $w_0$ and
$w_a$ when just three of these four parameters are allowed to vary, using data
from the CMB, SNe, and BAO,  
corresponding to $w_0 \simeq -1 \pm 0.2$, 
$w_a \sim 0\pm 1$ \citep{Kowalski_SCP}.
{\em While
the extant data are fully consistent with $\Lambda$CDM, they do not exclude
more exotic models of dark energy in which the dark energy density or its
equation-of-state parameter vary with time.}

\section{UNDERSTANDING COSMIC ACCELERATION}
\label{theory}
Understanding the origin of cosmic acceleration presents 
a stunning opportunity for theorists. 
As discussed in \S \ref{cosmology}, a smooth component with large negative 
pressure has repulsive gravity and can lead to the observed 
accelerated expansion within the context of GR. 
This serves to define dark energy. There is no shortage of ideas for 
what dark energy might be, from
the quantum vacuum to a new, ultra-light scalar field.
Alternatively, cosmic acceleration may arise from new gravitational 
physics, perhaps involving extra spatial dimensions. 
Here, we briefly review the theoretical landscape.

\bigskip

\subsection{Dark energy models}\label{sec:DE_models}

\subsubsection{Vacuum energy}  \label{sec:vacuum}
Vacuum energy is simultaneously the most plausible and most puzzling 
dark energy candidate.
General covariance
requires that the stress-energy of the vacuum takes 
the form of a constant times
the metric tensor, $T_{\rm VAC}^{\mu\nu} = \rho_{\rm VAC} g^{\mu\nu}$. Because the
diagonal terms $(T^0_0, T^i_i)$ of the stress-energy tensor $T^\mu_\nu$ are the energy density
and minus the pressure of the fluid, and $g^\mu_\nu$ is just the Kronecker
delta, the vacuum has a pressure equal to minus its energy density, $p_{\rm
VAC}=-\rho_{\rm VAC}$.  This also means that vacuum energy is mathematically
equivalent to a cosmological constant.

Attempts to compute the value of the vacuum energy density 
lead to very large or divergent results.  For each mode of a quantum field 
there is a zero-point energy $\hbar \omega
/2$, so that the energy density of the quantum vacuum is given by 
\begin{equation}
\rho_{\rm VAC} = {1\over 2}\sum_{\rm fields}g_i\int_0^\infty 
\sqrt{k^2 + m^2}\,{d^3k\over (2\pi )^3} \simeq  \sum_{\rm fields} {g_i k_{\rm max}^4\over 16\pi^2}
\end{equation}
where $g_i$ accounts for the degrees of freedom of the field (the sign of
$g_i$ is $+$ for bosons and $-$ for fermions), and the sum runs over all
quantum fields (quarks, leptons, gauge fields, etc). Here $k_{\rm max}$ is an
imposed momentum cutoff, because the sum diverges quartically.

To illustrate the magnitude of the problem, if the energy density contributed
by just one field is to be at most the critical density, then the cutoff
$k_{\rm max}$ must be $< 0.01\,{\rm eV}$ --- well below any energy scale where
one could have appealed to ignorance of physics beyond. [Pauli apparently
carried out this calculation in the 1930's, using the electron mass scale for
$k_{\rm max}$ and finding that the size of the Universe, that is, $H^{-1}$, 
 ``could not even reach
to the moon'' \citep{Straumann}.]  Taking the cutoff to be the Planck scale
($\approx 10^{19}$\,GeV), where one expects quantum field theory in a classical
spacetime metric to break down, the zero-point energy density would exceed the
critical density by some 120 orders-of-magnitude! It is very unlikely that 
a classical contribution to the vacuum energy density would cancel this 
quantum contribution to such high precision. This very large discrepancy
is known as the cosmological constant problem \citep{WeinbergRMP}. 

Supersymmetry, the hypothetical symmetry between bosons and fermions, appears 
to provide only partial help.
In a supersymmetric (SUSY) world, every fermion in the standard
model of particle physics has an equal-mass SUSY bosonic partner and vice
versa, so that fermionic and bosonic zero-point contributions to $\rho_{\rm
VAC}$ would exactly cancel.  However, SUSY is not a manifest symmetry in
Nature: none of the SUSY particles has yet been observed in collider
experiments, so they must be substantially heavier than their standard-model
partners. If SUSY is spontaneously broken at a mass scale $M$, one expects
the imperfect cancellations to generate a finite vacuum energy density
$\rho_{\rm VAC} \sim M^4$.  For the currently favored value $M \sim 1$ TeV,
this leads to a discrepancy of 60 (as opposed to 120) orders of magnitude with
observations. 
Nonetheless,  
experiments at the Large Hadron Collider (LHC) at CERN will
soon begin searching for signs of SUSY, e.g., SUSY partners of the quarks and
leptons, and might shed light on the vacuum-energy problem.

Another approach to the cosmological constant problem involves the idea that
the vacuum energy scale is a random variable that can take on
different values in different disconnected regions of the Universe.  Because a
value much larger than that needed to explain the observed cosmic acceleration
would preclude the formation of galaxies (assuming all other cosmological
parameters are held fixed), we could not find ourselves in a region with such
large $\rho_{\rm VAC}$ \citep{Weinberg87}.  This anthropic approach finds a
possible home in the landscape version of string theory, in which the number of
different vacuum states is very large and essentially all values of the
cosmological constant are possible.  Provided that the Universe has such a
multiverse structure, this might provide an explanation for the smallness of
the cosmological constant \citep{Bousso_Polchinski,Susskind}.

\subsubsection{Scalar fields}\label{sec:scalar_fields}

\begin{figure}[!t]
\centerline{\includegraphics[height=2.5in, width= 3in]{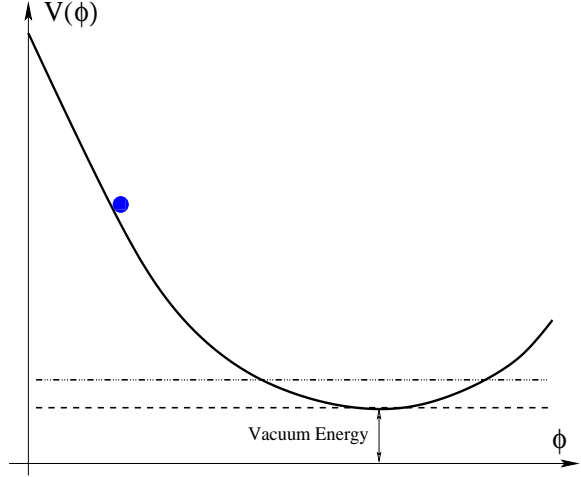}}
\caption{Generic scalar potential $V(\phi)$.  The scalar field rolls down 
the potential eventually settling at its minimum, which corresponds to the
vacuum.  The energy associated with the vacuum can be positive, negative, or
zero.}
\label{fig:Vphi}
\end{figure}

Vacuum energy does not vary with space or time and is not dynamical.  However,
by introducing a new degree of freedom, a scalar field $\phi$, one can make
vacuum energy effectively dynamical 
\citep{Ratra_Peebles,Wetterich,Frieman_PNGB,Zlatev}.  For a
scalar field $\phi$, with Lagrangian density ${\cal L} = {1\over
2}\partial^\mu\phi\partial_\mu\phi - V(\phi )$, the stress-energy takes the
form of a perfect fluid, with
\begin{equation}
\rho = \dot{\phi}^2/2  + V(\phi ) ~~,~~~~ p    = \dot{\phi}^2/2  - V(\phi )~,
\label{field_pot_kin}
\end{equation}
where $\phi$ is assumed to be spatially homogeneous, i.e., $\phi({\vec
x},t)=\phi(t)$, $\dot\phi^2/2$ is the kinetic energy, and $V(\phi )$ is the
potential energy; see Fig.~\ref{fig:Vphi}. The evolution of the field is
governed by its equation of motion,
\begin{equation}
  \ddot {\phi} + 3 H\dot{\phi} + V'(\phi) = 0~, 
\label{eq:field_roll}
\end{equation}
where a prime denotes differentiation with respect to $\phi$. 
Scalar-field dark energy can be described by the equation-of-state parameter
\begin{equation}
w = {\dot\phi^2/2 - V(\phi ) \over \dot\phi^2/2 + V(\phi )} =
{-1 + \dot\phi^2/2V \over 1 + \dot\phi^2/2V}~.
\label{eq:wphi}
\end{equation}
If the scalar field evolves slowly, $\dot{\phi}^2/2V \ll 1$, then $w\approx
-1$, and the scalar field behaves like a slowly varying vacuum energy, with
$\rho_{\rm VAC}(t) \simeq V[\phi (t)]$.  In general, from Eq. (\ref{eq:wphi}),
$w$ can take on any value between $-1$ (rolling very slowly) and $+1$
(evolving very rapidly) and varies with time.

Many scalar field models can be classified dynamically 
as ``thawing'' or ``freezing'' \citep{Caldwell_Linder}. In freezing models, the field 
rolls more slowly as time progresses, i.e., the slope of the potential drops 
more rapidly than the Hubble friction term $3H\dot\phi$ in \ Eq.~(\ref{eq:field_roll}).   
This can happen if, e.g., $V(\phi)$ falls off exponentially or as an inverse power-law 
at large $\phi$. For thawing models, at early times the field is frozen by the 
friction term, and it acts as vacuum energy; when the expansion rate drops 
below $H^2 = V''(\phi)$, the field begins to roll and $w$ evolves away from $-1$. 
The simplest example of a thawing model is a 
scalar field of mass $m_\phi$, with $V(\phi)=m^2_\phi \phi^2/2$.
Since thawing and freezing fields tend to 
have different trajectories of $w(z)$, 
precise cosmological measurements might be able to discriminate 
between them.

\subsubsection{Cosmic coincidence and Scalar Fields}

As Fig.~\ref{fig:scalings} shows, through most of the history of the Universe,
dark matter or radiation dominated dark energy by many orders of magnitude.
We happen to live around the time that dark energy has become important.  Is
this coincidence between $\rho_{\rm DE}$ and $\rho_{\rm M}$ an important clue to
understanding cosmic acceleration or just a natural consequence of the
different scalings of cosmic energy densities and the longevity of the
Universe? In some freezing models, the scalar field energy density tracks that
of the dominant component (radiation or matter) at early times and then
dominates at late times, providing a dynamical origin for the coincidence. 
In thawing models, the coincidence is indeed
transitory and just reflects the mass scale of the scalar field.

\subsubsection{More complicated scalar-field models}
While the choice of the potential $V(\phi)$ allows a large range of dynamical
behaviors, theorists have also considered the implications of modifying the
canonical form of the kinetic energy term ${1\over
2}\partial^\mu\phi\partial_\mu\phi$ in the Lagrangian.  By changing the sign of
this term, from Eq. \ref{eq:wphi} it is possible to have $w < -1$
\citep{Caldwell_phantom}, although such theories are typically unstable
\citep{CHT03}.  In ``k-essence,'' one introduces a field-dependent kinetic term
in the Lagrangian to address the coincidence problem \citep{kessence_1}.

\subsubsection{Scalar-field issues}
Scalar-field models raise new questions and possibilities.  For example, is
cosmic acceleration related to inflation?  After all, both involve accelerated
expansion and can be explained by scalar field dynamics.  Is dark energy
related to dark matter or neutrino mass?  No firm or compelling connections
have been made to either, although the possibilities are intriguing. Unlike
vacuum energy, which must be spatially uniform, scalar-field dark energy can
clump, providing a possible new observational feature, but in most cases is
only expected to do so on the largest observable scales today (see \S 10.2.1).

Introducing a new dynamical degree of freedom allows for a richer variety of
explanations for cosmic acceleration, but it is not a panacea.  Scalar field
models do not address the cosmological constant problem: they simply assume
that the minimum value of $V(\phi)$ is very small or zero; see Fig. \ref{fig:Vphi}. 
Cosmic acceleration is then attributable to the fact that the Universe 
has not yet reached its true vacuum state, for dynamical reasons. 
These models also pose new
challenges: in order to roll slowly enough to produce accelerated expansion,
the effective mass of the scalar field must be very light compared to other
mass scales in particle physics, $m_\phi \equiv \sqrt{V''(\phi)} \lesssim 3H_0
\simeq 10^{-42}$ GeV, even though the field amplitude is typically of order
the Planck scale, $\phi \sim 10^{19}$ GeV.  This hierarchy, $m_\phi/\phi \sim
10^{-60}$, means that the scalar field potential must be extremely flat.
Moreover, in order not to spoil this flatness, the interaction strength of the
field with itself must be extremely weak, at most of order $10^{-120}$ in
dimensionless units; its coupling to matter must also be very weak to be
consistent with constraints upon new long-range forces
\citep{Carroll_quint}. Understanding such very small numbers and ratios makes
it challenging to connect scalar field dark energy with particle physics
models \citep{Frieman_PNGB}. In constructing theories that go beyond the standard model of particle
physics, including those that incorporate primordial inflation, model-builders
have been strongly guided by the requirement that any small dimensionless
numbers in the theory should be protected by symmetries from large quantum
corrections (as in the SUSY example above). Thus far, such model-building
discipline has not been the rule among cosmologists working on dark energy
models.

\subsection{Modified gravity} \label{sec:MG}

A very different approach holds that cosmic acceleration is a manifestation of
new gravitational physics rather than dark energy, i.e., that it involves 
a modification of the geometric as opposed to the stress-tensor side of the 
Einstein equations.  Assuming that 4-d
spacetime can still be described by a metric, the operational changes are
twofold: (1) a new version of the Friedmann equation governing the evolution
of $a(t)$; (2) modifications to the equations that govern the growth of the
density perturbations that evolve into large-scale structure.  A number of
ideas have been explored along these lines, from models motivated by
higher-dimensional theories and string theory
\citep{DGP,Deffayet}
to phenomenological modifications of the
Einstein-Hilbert Lagrangian of GR \citep{CDTT,Song_Hu_fR}.

Changes to the Friedmann equation are easier to derive, discuss, and analyze.
In order not to spoil the success of the standard cosmology at early times
(from big bang nucleosynthesis to the CMB anisotropy to the formation of
structure), the Friedmann equation must reduce to the GR form for $z \gg 1$.
As a specific example, consider the model of \citet{DGP}, which arises
from a five-dimensional gravity theory and has a 4-d Friedmann equation,
\begin{equation}
  H^2 = {8\pi G \rho \over 3} + {H\over r_c}~,
\end{equation}
where $r_c$ is a length scale related to the 5-dimensional
gravitational constant.  As the energy density in matter and radiation, $\rho$,
becomes small, there is an accelerating solution, with $H= 1/r_c$. From the
viewpoint of expansion, the additional term in the Friedmann equation has the
same effect as dark energy that has an equation-of-state parameter which
evolves from $w=-1/2$ (for $z \gg 1$) to $w = -1$ in the distant future. 
While attractive, it is not clear that a consistent model with 
this dynamical behavior exists \citep[e.g.,][]{Gregory07}.

\subsection{Unmodified gravity}

Instead of modifying the right or left side of the Einstein equations to 
explain the supernova observations, a third logical possibility is to 
drop the assumption that the Universe is spatially homogeneous on large scales. 
It has been argued that the non-linear gravitational effects of spatial 
density perturbations, when averaged over large scales, could yield 
a distance-redshift relation in our observable patch of the Universe 
that is very similar to that for an accelerating, homogeneous 
Universe \citep{Kolb_Matarrese_06}, obviating the need for either dark energy 
or modified gravity. While there has been 
debate about the amplitude of these effects, this idea has helped 
spark renewed interest in a class of exact, inhomogeneous cosmologies. 
For such Lema\^{\i}tre-Tolman-Bondi models
to be consistent with the SN data and
not conflict with the isotropy of the CMB, the Milky Way must be near the
center of a very large-scale, nearly spherical, underdense region
\citep{Tomita_00,Alnes_05,Enqvist_07}.  
Whether or not such models can be made consistent with the wealth of precision
cosmological data remains to be seen; moreover, requiring our galaxy 
to occupy a privileged location, in violation of the spirit of the Copernican 
principle, is not yet theoretically well-motivated.

\subsection{Theory summary}
There
is no compelling explanation for cosmic acceleration, but 
many intriguing ideas are being explored.  Here is our assessment:

\begin{itemize}
\item Cosmological constant: Simple, but no underlying physics 
\item Vacuum energy: Well-motivated, mathematically equivalent to a cosmological 
constant; $w=-1$ is consistent with all data, but all
  attempts to estimate its size are at best orders of magnitude too large 
\item Scalar fields: Temporary period of cosmic acceleration, $w$ varies
  between $-1$ and $1$ (and could also be $<-1$), possibly related to inflation, but
  does not address the cosmological constant problem and may lead to new long-range forces 
\item New gravitational physics: Cosmic acceleration could be a clue to going beyond GR, but no
  self-consistent model has been put forth 
\item Old gravitational physics:  It may be possible to find an 
inhomogeneous solution that is observationally viable, but such solutions do not yet seem compelling
\end{itemize}

The ideas underlying many of these approaches, 
from attempting to explain the smallness of quantum vacuum energy to extending Einstein's
theory, are bold.  Solving the puzzle of cosmic acceleration thus has
the potential to advance our understanding of many important problems in
fundamental physics.

\section{DESCRIBING DARK ENERGY}
\label{describingDE}
The absence of a consensus model for cosmic acceleration 
presents a challenge in trying
to connect theory with observations.  For dark energy, the
equation-of-state parameter 
$w$ provides a useful phenomenological description \citep{Turner_White}.
Because it is the ratio of pressure to energy density, it is also 
closely connected
to the underlying physics.  However, $w$ is not fundamentally a function of
redshift, and if cosmic acceleration is due to new gravitational physics, the
motivation for a description in terms of $w$ disappears.  
In this section, we review the variety of formalisms that have
been used to describe and constrain dark energy.

\subsection{Parametrizations} \label{sec:Taylor}

The simplest parameterization of dark energy is $w={\rm const}$.  This form
fully describes vacuum energy ($w=-1$) 
and, together with $\ode$ and $\Omega_{\rm M}$, 
provides a 3-parameter description of the
dark-energy sector (2 parameters if flatness is assumed). 
However, it does not describe scalar 
field or modified gravity models.

A number of two-parameter descriptions of $w$ have been explored, e.g., $w(z) =
w_0 + w' z$ and $w(z) = w_0 + b\ln (1+z)$.
For low redshift they are all essentially equivalent, but for large $z$, some
lead to unrealistic behavior, e.g., $w\ll -1$ or $\gg 1$.  The
parametrization 
\begin{equation}
w(a)=w_0 + w_a(1-a)= w_0+w_a z/(1+z) 
\label{eq:w0wa}
\end{equation}
\citep[e.g.,][]{Linder_wa} avoids this problem and leads to the most commonly used description
of dark energy, namely $(\ode,\Omega_{\rm M},w_0,w_a)$ 

More general expressions have been proposed, for example, Pad\'{e} approximants
or the transition between two asymptotic values $w_0$ (at $z\rightarrow 0$) and
$w_f$ (at $z\rightarrow \infty$), $w(z) = w_0 + (w_f-w_0)/(1+
\exp[(z-z_t)/\Delta ])$ \citep{Corasaniti_Copeland}.

\begin{figure}[!t]
\begin{center}
\centerline{\psfig{file=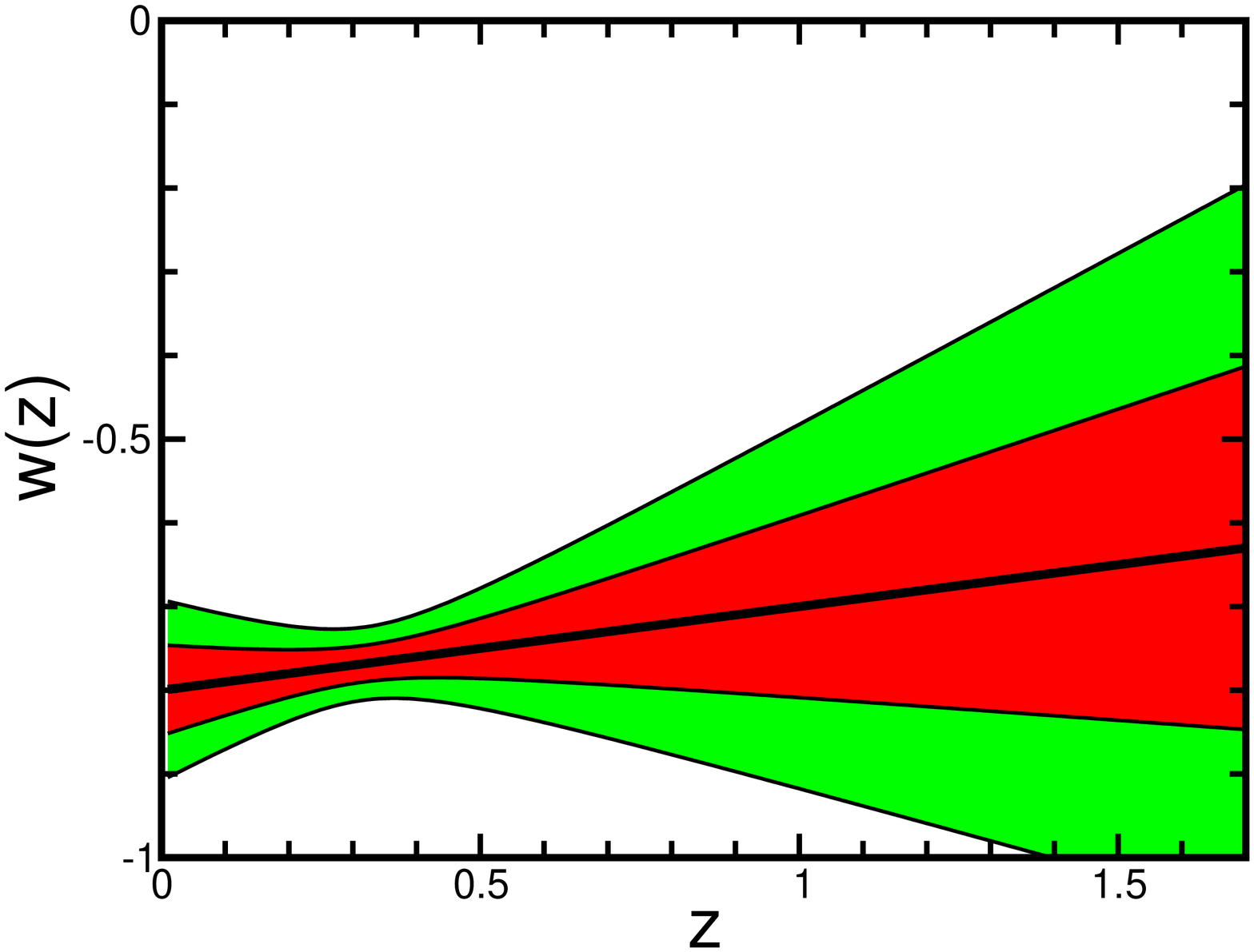,width=3.3in,height=2.8in,angle=-90}
\hspace{-1.3cm}
\vspace{-0.5cm}
\psfig{file=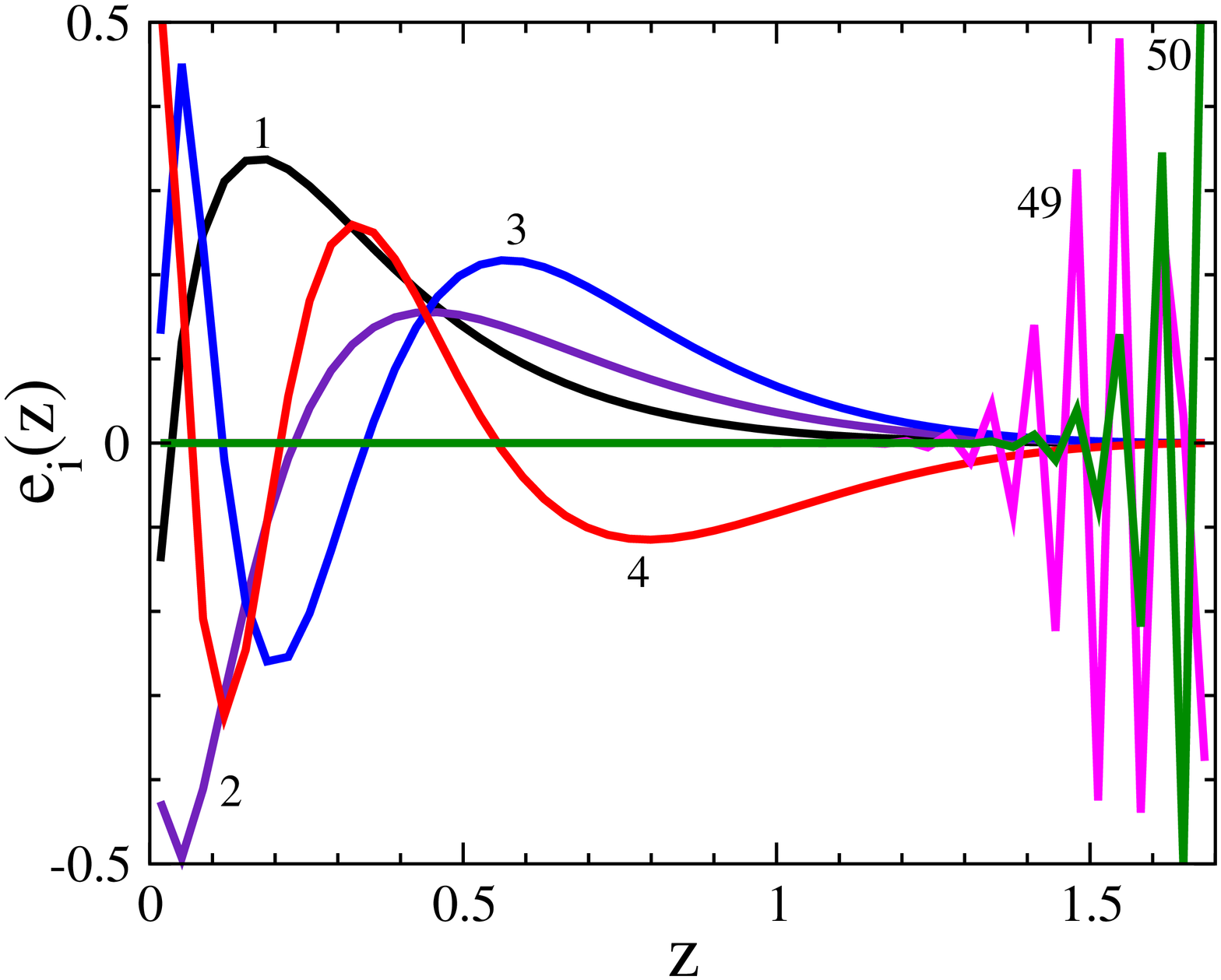,width=3.2in,height=2.5in,angle=-90}}
\caption{{\it Left panel:} Example of forecast 
constraints on $w(z)$, assuming $w(z)=w_0+w'z$.  The ``pivot'' redshift,
$z_p\simeq 0.3$, is where $w(z)$ is best determined. From
\citet{Hut_Tur_00}. {\it Right panel:} The four best-determined (labelled
$1-4$) and two worst-determined (labelled $49, 50$) principal components of
$w(z)$ for a future SN Ia survey such as SNAP, with several thousand SNe in the
redshift range $z=0$ to $z=1.7$. From \citet{Huterer_Starkman}.  }
\label{fig:w0wp_PC}
\end{center}
\end{figure}

The two-parameter descriptions of $w(z)$ that are linear in the parameters
entail the existence of a ``pivot'' redshift $z_p$ at which the measurements of
the two parameters are uncorrelated and the error in $w_p \equiv w(z_p)$
reaches a minimum
\citep{Hut_Tur_00}; see the left panel of Fig.~\ref{fig:w0wp_PC}.  
The redshift of this sweet spot varies with the cosmological probe and survey
specifications; for example, for current SN Ia surveys $z_p\approx 0.25$.  Note
that forecast constraints for a particular experiment on $w_p$ are numerically
equivalent to constraints one would derive on constant $w$.

\subsection{Direct reconstruction}\label{sec:reconstruction}

Another approach is to directly invert the redshift-distance relation 
$r(z)$ measured 
from SN data to obtain the redshift dependence of $w(z)$ in terms 
of the first and second derivatives of the comoving distance
\citep{reconstr,
Starobinsky},
\begin{equation}
1+w(z) = {1+z\over 3}\, {3H_0^2\Omega_{\rm M}(1+z)^2 + 2(d^2r/dz^2)/(dr/dz)^3\over
        H_0^2\Omega_{\rm M}(1+z)^3-(dr/dz)^{-2}}~.
\label{eq:wz_reconstr}
\end{equation}
Assuming that dark energy is due to a single rolling scalar field, the scalar
potential can also be reconstructed,
\begin{eqnarray}
V[\phi (z)] 
&=& {1\over 8\pi G}\left[ {3\over (dr/dz)^2}
        +(1+z) {d^2r/dz^2\over (dr/dz)^3}\right]  - 
	{3\Omega_{\rm M}H_0^2  (1+z)^3 \over 16\pi G} 
\label{eq:Vphi_reconstr}
\end{eqnarray}
%%
%where the upper (lower) sign applies if $\dot\phi >0$ ($<0$).  
Others have
suggested reconstructing the dark energy density
\citep{Wang_Mukherjee_03}, 
\begin{equation}
\rho_{\rm DE}(z) = {3\over 8\pi G} \left[{1\over (dr/dz)^2} - \Omega_{\rm M}H_0^2 (1+z)^3 \right] \, .
\end{equation}

Direct reconstruction is the only approach that is truly
model-independent. However, it comes at a price -- taking 
derivatives of noisy data.  In practice, one must fit the distance data with a
smooth function --- e.g., a polynomial, Pad\'{e} approximant, or spline with
tension, and the fitting process introduces systematic biases.  While a variety
of methods have been pursued
\citep[e.g.,][]{
Weller_Albrecht,
Gerke_Efstat},
it appears that direct reconstruction is too
challenging and not robust even with SN Ia data of excellent quality.  
Although the expression for $\rho_{\rm DE}(z)$ involves only 
first derivatives of $r(z)$, it contains little information about 
the nature of dark energy. 
For a
review of dark energy reconstruction and related issues, see
\citet{Sahni_review}.

\subsection{Principal components}\label{sec:PC}

The cosmological function that we are trying to
determine --- $w(z)$, $\rho_{\rm DE}(z)$, or $H(z)$ --- can be 
expanded in terms of principal components, a set of functions 
that are uncorrelated and orthogonal by construction  
\citep{Huterer_Starkman}. In this approach, the data
determine which components are measured best.

For example, suppose we parametrize $w(z)$
in terms of piecewise constant values $w_i$ ($i=1, \ldots, N$), each defined
over a small redshift range ($z_i$, $z_i+\Delta z$).  In the limit of small
$\Delta z$ this recovers the shape of an arbitrary dark energy history (in
practice, $N\gtrsim 20$ is sufficient), but the estimates of the 
$w_i$ from a given dark energy probe will be very noisy for large $N$. 
Principal Component Analysis extracts from those noisy estimates 
the best-measured features of $w(z)$. We find the eigenvectors $e_i(z)$ of
the inverse covariance matrix for the parameters $w_i$ and the corresponding
eigenvalues $\lambda_i$. The equation-of-state parameter is then expressed as
\begin{equation}
w(z) = \sum_{i=1}^N \alpha_i\, e_i(z)~,
\label{eq:w_expand}
\end{equation}
where the $e_i(z)$ are the principal components. The coefficients 
$\alpha_i$, which can be computed via the orthonormality condition, 
are each determined with
an accuracy $1/\sqrt{\lambda_i}$. Several of these components are
shown for a future SN survey in the right panel of Fig. ~\ref{fig:w0wp_PC}.

One can use this approach to design a survey that is most sensitive
to the dark energy equation-of-state parameter in some 
specific redshift interval or to study how
many independent parameters are measured well by a combination of
cosmological probes.
There are a variety of extensions of
this method, including measurements of the equation-of-state 
parameter in redshift intervals 
\citep{Huterer_Cooray}.

\subsection{Kinematic description}\label{sec:kinematic}

If the explanation of cosmic acceleration is a modification of GR and not dark
energy, then a purely kinematic description through, e.g., the
functions $a(t)$, $H(z)$, or $q(z)$ may be the best approach. With the weaker
assumption that gravity is described by a metric theory and that spacetime is
isotropic and homogeneous, the FRW metric is still valid, as are the kinematic
equations for redshift/scale factor, age, $r(z)$, and volume element.  The
dynamical equations, i.e., the Friedmann equations and the growth of density
perturbations, may however be different.

If $H(z)$ is chosen as the kinematic variable, then $r(z)$ and age take their standard forms.  On the
other hand, to describe acceleration one might wish to take the deceleration
parameter $q(z)$ as the fundamental variable; the expansion rate is then given
by
\begin{equation}
H(z) = H_0 \exp \left[ \int_0^z [1 + q(z^\prime )]d\ln (1+z^\prime ) \right].
\label{eq:qH}
\end{equation}

Another possibility is the dimensionless ``jerk'' parameter, $j\equiv
(\dddot{a}/a)/H^3$, instead of $q(z)$ \citep{Visser,Rapetti_Blandford}.
The deceleration $q(z)$ can be
expressed in terms of $j(z)$,
\begin{equation}
{dq\over d\ln (1+z)} + q(2q+1)-j=0 ~,
\label{eq:jerk}
\end{equation}
and, supplemented by Eq.~(\ref{eq:qH}), $H(z)$ may be obtained. 
Jerk has the virtue that 
constant $j=1$ corresponds to
a cosmology that transitions from $a \propto t^{2/3}$ at early times to $a
\propto e^{Ht}$ at late times. Moreover, for constant jerk,  
Eq.~(\ref{eq:jerk}) is easily solved: 
\begin{equation}
\ln \left[ {q-q_+ \over q-q_-} \right]   = \exp \left[ -2(q_+ - q_-)(1+z) \right]\,,
\qquad q_\pm =  {1\over 4}(-1\pm\sqrt{1+8j}) ~.
\end{equation}
On the other hand, constant jerk does not span cosmology model-space well: the
asymptotic values of deceleration are $q = q_\pm$, so that only for $j=1$ can
there be a matter-dominated beginning ($q={1\over 2}$).  
One would test for departures from $\Lambda$CDM by searching for 
variation of $j(z)$ from unity over some redshift interval; in 
principle, the same information is also encoded in $q(z)$. 

The kinematic approach has produced some interesting results;
using the SN data and the principal component method, \citet{Shapiro_Turner}
find the best measured mode of $q(z)$ can be used to infer 5-$\sigma$ evidence
for acceleration of the Universe at {\em some} recent time, without recourse to
GR and the Friedmann equation.

\section{PROBES OF COSMIC ACCELERATION}
\label{probes}
As described in \S \ref{currentstatus}, the phenomenon of accelerated
expansion is now well established, and the dark energy density 
has been determined to a precision of a few percent. However, getting at 
the {\it nature} of the dark energy---by measuring its 
equation-of-state parameter---is more challenging. 
To illustrate, consider that 
for fixed $\Omega_{\rm DE}$, a 1\% change in (constant)
$w$ translates to only a 3\% (0.3\%) change in dark-energy (total) density at redshift 
$z=2$ and only 
a 0.2\% change in distances to redshifts $z=1-2$.

The primary effect of dark energy is on the expansion rate of the
Universe; in turn, this affects the redshift-distance relation 
and the growth of structure.
While dark energy has been important at recent epochs, 
we expect that its effects
at high redshift were very small, since otherwise it would have been 
difficult for large-scale structure to have formed (in most models). 
Since 
$\rho_{\rm DE}/\rho_{\rm M} \propto (1+z)^{3w}
\sim 1/(1+z)^3$, the redshifts of highest leverage 
for probing dark energy are expected to be 
between a few tenths and two \citep{Hut_Tur_00}. Four methods hold particular
promise in probing dark energy in this redshift range: type Ia supernovae,
clusters of galaxies, baryon acoustic oscillations, and weak gravitational
lensing.  In this section, we describe and compare these four probes,
highlighting their complementarity in terms of both dark energy constraints
and the systematic errors to which they are susceptible.  Because of this
complementarity, a multi-pronged approach will be most effective. The goals 
of the next generation of dark energy experiments, described in \S \ref{projects}, 
are to constrain $w_0$ at the few percent level and $w_a$ at the 10\% level. 

While our focus is on these four techniques, we also briefly discuss other
dark-energy probes, emphasizing the important supporting role of the CMB.  

\subsection{Supernovae}\label{sec:SN}

By providing bright, standardizable candles \citep{Leibundgut_01}, type Ia supernovae constrain cosmic
acceleration through the Hubble diagram, cf., Eq. (\ref{eq:distmod}).  
The first
direct evidence for cosmic acceleration came from SNe Ia, and they have 
provided the strongest constraints on the dark energy equation-of-state 
parameter.
At present, they are the most effective and mature
probe of dark energy.

\begin{figure}[!t]
\centering
\includegraphics[trim = 0mm 20mm 0mm 10mm, clip, width=3.in]{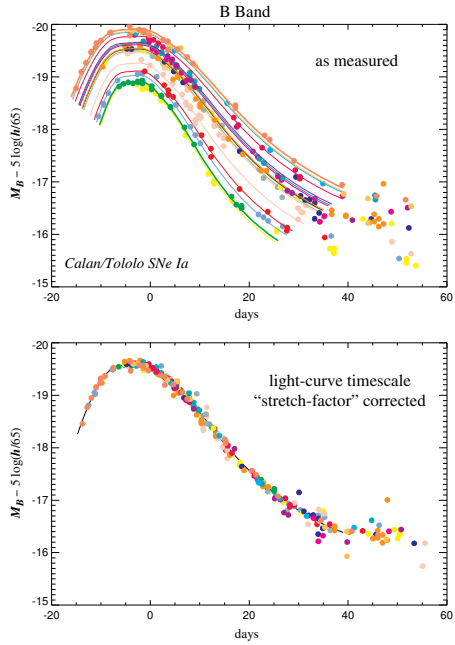}
\caption{
{\it Top panel:} $B$-band light curves for low-redshift SNe Ia from the Calan-Tololo 
survey \citep{hamuy_96} show an intrinsic scatter of $\sim 0.3$ mag in peak 
luminosity. 
{\it Bottom panel:} 
After a one-parameter correction for the brightness-decline correlation, the light
curves show an intrinsic dispersion of only $\sim 0.15\,$mag. From \citet{Kim_04}.  }
\label{fig:stretch}
\end{figure}

SN Ia light curves are powered by the radioactive decays of ${}^{56}$Ni (at
early times) and ${}^{56}$Co (after a few weeks), produced in the
thermonuclear explosion of a carbon-oxygen white dwarf accreting mass from a
companion star as it approaches the Chandrasekhar mass
\citep{Hillebrandt_00}.  The peak luminosity is determined by the
mass of ${}^{56}$Ni produced in the explosion \citep{Arnett_82}: if the white
dwarf is fully burned, one expects $\sim 0.6 M_\odot$ of ${}^{56}$Ni to be
produced.  As a result, although the detailed mechanism of SN Ia explosions 
remains uncertain \citep[e.g.,][]{hoflich04,plewa},
SNe Ia are expected to have
similar peak luminosities.  Since they are about as bright as a typical galaxy
when they peak, SNe Ia can be observed to large distances, recommending their
utility as standard candles for cosmology.

In fact, as Fig.~\ref{fig:stretch} shows, SNe Ia are not intrinsically 
standard candles, with a $1\sigma$ spread of
order 0.3 mag in peak $B$-band luminosity which would limit
their utility.  However, work in the early 1990's \citep{Phillips_93}
established an empirical correlation between SN Ia peak brightness and the
rate at which the luminosity declines with time after peak: intrinsically
brighter SNe Ia decline more slowly. After correcting for this correlation,
SNe Ia turn out to be excellent ``standardizable'' candles, with a dispersion
of about 15\% in peak brightness.

Cosmological
parameters are constrained by comparing distances to low- and high-redshift
SNe Ia.  Operationally, since $H_0d_L$ is independent of the Hubble parameter
$H_0$, Eq.~(\ref{eq:distmod}) can be written as $m= 5\log_{10}[H_0
d_L(z;\Omega_{\rm M},\ode,w(z))]+\cal{M}$, where ${\cal{M}} \equiv 
M-5\log_{10}(H_0 ~{\rm Mpc})+25$ is the parameter effectively constrained by the 
low-redshift SNe that anchor the Hubble diagram.

The major systematic concerns for supernova distance measurements are errors
in correcting for host-galaxy extinction and uncertainty in the intrinsic colors
of SNe Ia; luminosity evolution; and selection bias in the low-redshift sample.  
For observations in two passbands, with perfect knowledge of
intrinsic SN colors or of the extinction law, one could solve for the
extinction and eliminate its effects on the distance modulus.  In practice, the
combination of photometric errors, variations in intrinsic SN colors, and
uncertainties and likely variations in host-galaxy dust properties lead to
distance uncertainties even for multi-band observations of SNe. Observations
that extend into the rest-frame near-infrared, where the
effects of extinction are much reduced, offer promise in controlling 
this systematic. 

With respect to luminosity evolution, there is evidence 
that SN peak luminosity correlates with
host-galaxy type \citep[e.g.,][]{JRK_07}, and that the mean host-galaxy environment,
e.g., the star formation rate, evolves strongly with look-back time.  However,
brightness-decline-corrected SN Ia Hubble diagrams are consistent between
different galaxy types, and since the nearby Universe spans the range of
galactic environments sampled by the high-redshift SNe, one can measure
distances to high-redshift events by comparing with low-redshift analogs.
While SNe provide a number of correlated observables (multi-band light curves
and multi-epoch spectra) to constrain the physical state of the system,
insights from SN Ia theory will likely be needed to determine if they are collectively
sufficient to constrain the mean peak luminosity at the percent level
\citep{hoflich04}.  

Finally, there is concern that the low-redshift SNe currently used to anchor 
the Hubble diagram and that serve as templates for fitting
distant SN light curves are a relatively small, heterogeneously selected
sample and that correlated large-scale peculiar velocities induce 
larger distance errors than previously estimated \citep{hui_06}.  
This situation should improve in the near future once results are
collected from low-redshift SN surveys such as the Lick Observatory Supernova
Search (LOSS), the Center for Astrophysics Supernova project, the Carnegie Supernova
Project, the Nearby Supernova Factory, and the Sloan Digital Sky Survey-II
Supernova Survey.

Accounting for systematic errors, precision measurement of $w_0$ and 
$w_a$ with SNe will require a few thousand 
SN Ia light curves out to redshifts $z\sim 1.5$ to be measured with
unprecedented precision and control of systematics \citep{frieman_03}.  
For redshifts $z>0.8$, this will 
require going to space to minimize photometric errors, to obtain uniform
light-curve coverage, and to observe in the near-infrared bands to capture the 
redshifted photons.

\subsection{Clusters}\label{sec:clusters}

Galaxy clusters are the largest virialized objects in the Universe.
Within the context of the CDM paradigm, the number density of
cluster-sized dark matter halos as a function of redshift and halo mass can be
accurately predicted from N-body simulations \citep{warren_06}. Comparing
these predictions to
large-area cluster surveys that extend to high redshift ($z \gtrsim 1$) can
provide precise constraints on the cosmic expansion history
\citep{wang_98,haiman_01}. 

The redshift distribution of clusters in a survey that selects clusters
according to some observable $O$ with redshift-dependent selection function
$f(O,z)$ is given by
\begin{equation}
\frac{d^{2}N(z)}{dzd\Omega} = \frac{r^2(z)}{H(z)} 
\int^{\infty}_{0}f(O,z)dO\int^{\infty}_{0}p(O|M,z)\frac{dn(z)}{dM}dM ~,
\label{eq:clustercount}
\end{equation}
where $dn(z)/dM$ is the space density of dark halos in comoving coordinates,
and $p(O|M,z)$ is the mass-observable relation, the probability that a
halo of mass $M$ at redshift $z$ is observed as a cluster with observable
property $O$. The utility of this  probe hinges on the ability 
to robustly associate cluster observables such as X-ray luminosity or
temperature, cluster galaxy richness, Sunyaev-Zel'dovich effect flux decrement,
or weak lensing shear, with cluster mass
\citep[e.g.,][]{borgani_06}.

The sensitivity of cluster counts to dark energy arises from two factors: {\it
geometry}, the term multiplying the integral in Eq.~(\ref{eq:clustercount})
is the comoving volume element; and {\it growth of structure}, $dn(z)/dM$
depends on the evolution of density perturbations, cf. Eq.~\ref{eq:growth}. 
The cluster mass function is also
determined by the primordial spectrum of density perturbations; its
near-exponential dependence upon mass is the root of the power of 
clusters to probe dark energy.

\begin{figure}[!t]
\centerline{
\psfig{file=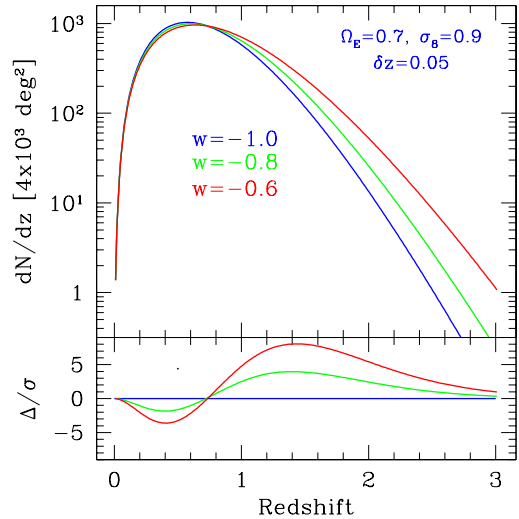,width=3.in}
}
\caption{ 
Predicted cluster counts for a survey covering 4,000 sq.\ deg.\ 
that is sensitive to halos more massive than $2\times 10^{14} M_\odot$, for 3 flat cosmological models with fixed $\Omega_{\rm M}=0.3$ and $\sigma_8=0.9$. 
Lower panel shows differences between the models relative to the 
statistical errors. 
From \citet{Mohr_04}.
}
\label{fig:massfun}
\end{figure}

Fig.~\ref{fig:massfun} shows the sensitivity to the dark energy
equation-of-state parameter of the expected cluster counts for the South Pole Telescope
and the Dark Energy Survey. At modest redshift, $z<0.6$, the differences are
dominated by the volume element; at higher redshift, the counts are most
sensitive to the growth rate of perturbations.

The primary systematic concerns are uncertainties in the mass-observable
relation $p(O|M,z)$ and in the selection function $f(O,z)$.  The strongest
cosmological constraints arise for those cluster 
observables that are most strongly
correlated with mass, i.e., for which $p(O|M,z)$ is narrow for fixed $M$, 
and which have a
well-determined selection function.
There are several independent techniques both for detecting clusters and for
estimating their masses using observable proxies.  Future surveys will aim to
combine two or more of these techniques to cross-check cluster mass estimates
and thereby control systematic error. Measurement of the spatial correlations 
of clusters and of the shape of the mass function provide additional 
internal calibration of the mass-observable relation \citep{Majumdar_04,Lima_04}.

With multi-band CCD imaging, clusters can be efficiently detected as
enhancements in the surface density of early-type galaxies, and their observed
colors provide photometric 
redshift estimates that substantially reduce the projection
effects that plagued early optical cluster catalogs \citep{Yee_01,Koester_07}.
Weak lensing and dynamical studies show that cluster richness correlates with
cluster mass \citep{Johnston_07} and can be used to statistically calibrate 
mass-observable relations.  Most of the cluster baryons reside in hot,
X-ray emitting gas in approximate dynamical equilibrium in the dark matter
potential well.  Since X-ray luminosity is proportional to the square of the
gas density, X-ray clusters are high-contrast objects, for which the selection
function is generally well-determined. Empirically, X-ray luminosity and
temperature are both found to correlate more tightly than optical richness with
virial mass \citep{Arnaud_05,Stanek_Evrard}.  

The hot gas in clusters also Compton
scatters CMB photons as they pass through, leading to the Sunyaev-Zel'dovich
effect \citep[SZE;][]{SZ_70}, a measurable distortion of the blackbody
CMB spectrum.  It can
be detected for clusters out to high redshift \citep[e.g.,][]{Carlstrom_02}. 
Since the SZE flux decrement is linear in the gas density, it should be less
sensitive to gas dynamics \citep{Nagai_06,Motl_05}.  
Finally, weak gravitational lensing can be used both to
detect and to infer the masses of clusters. Since lensing is sensitive to all
mass along the line of sight, projection effects are the major concern for
shear-selected cluster samples \citep{Hennawi_Spergel,White_02}. 

X-ray or SZE measurements also enable measurements of the baryonic gas mass in
clusters; in combination with the virial mass estimates described above, this
enables estimates of the baryon gas fraction, $f_{\rm gas} \propto M_{\rm B}/M_{\rm
tot}$. The ratio inferred from X-ray/SZE measurements depends upon cosmological
distance because the inferred baryon mass, $M_{\rm B} \propto d_L^{5/2}$ (X-ray) or 
$\propto d_L^2$ (SZE), and the inferred total mass from X-ray measurements
$M_{\rm tot} \propto d_L$.  If clusters are representative samples of matter,
then $f_{\rm gas}(z) \propto d_L^{3/2~ {\rm or}~ 1}$ should be 
independent of redshift and $\approx
\Omega_{\rm B}/\Omega_{\rm M}$; this 
will only be true for the correct cosmology
\citep{Allen_07, Rapetti_07}.

\subsection{Baryon acoustic oscillations}
\label{sec:BAO}

The peaks and troughs seen in the angular power spectrum of the CMB 
temperature anisotropy (see Fig.~5) arise from gravity-driven acoustic 
oscillations of the coupled photon-baryon fluid in the early Universe. 
The scale of these oscillations is set by the sound horizon at the 
epoch of recombination---the distance $s$ that sound waves in the fluid 
could have traveled by that time, 
\begin{equation}
s= \int_{0}^{t_{rec}} c_s (1+z) dt = \int_{z_{rec}}^{\infty} {c_s \over H(z)}dz~,
\label{eq:sound}
\end{equation}
where the sound speed $c_s$ is determined by the ratio of the baryon and photon
energy densities. The precise measurement of the angular scales of the acoustic
peaks by WMAP has determined $s = 147\pm 2$\,Mpc.  After
recombination, the photons and baryons decouple, and the effective sound speed
of the baryons plummets due to the loss of photon pressure; the sound waves
remain imprinted in the baryon distribution and, through gravitational
interactions, in the dark matter distribution as well.  Since the sound horizon
scale provides a ``standard ruler'' calibrated by the CMB anisotropy,
measurement of the baryon acoustic oscillation (BAO) scale in the galaxy
distribution provides a geometric probe of the expansion history. 

In the galaxy power spectrum, this scale appears as a series of oscillations
with amplitude of order 10\%, more subtle than the acoustic oscillations in
the CMB because the impact of baryons on the far larger dark matter component
is small.  Measuring the BAO scale from galaxy clustering in the transverse
and line-of-sight directions yields measurements of $r(z)/s$ and of 
$sH(z)$, respectively \citep{Hu_rings,Seo_Eisenstein,Blake03}. 
Spectroscopic redshift surveys can probe both, while photometric surveys are
mainly sensitive to transverse clustering.  While determining these quantities
with precision requires enormous survey volumes and millions of galaxies,
N-body simulations suggest that the systematic uncertainties associated with BAO
distance scale measurements are smaller than those of other observational
probes of dark energy.  Because such large numbers of galaxies are needed, BAO
measurements provide distance estimates 
that are coarse-grained 
in redshift.

The main systematic uncertainties in the interpretation of BAO measurements
are the effects of non-linear gravitational evolution, of scale-dependent
differences between the clustering of galaxies and of dark matter (bias), and,
for spectroscopic surveys, redshift distortions of the clustering, which can shift the BAO features.
Numerical studies to date suggest that the resulting shift of the scale
of the BAO peak in the galaxy power spectrum is at the percent level or less 
\citep{Seo07, Guzik06, Smith07}, comparable to the forecast measurement 
uncertainty for future surveys but in principle predictable from
high-resolution simulations.

\subsection{Weak gravitational lensing}\label{sec:WL}

\begin{figure}[!t]
\centerline{\psfig{file=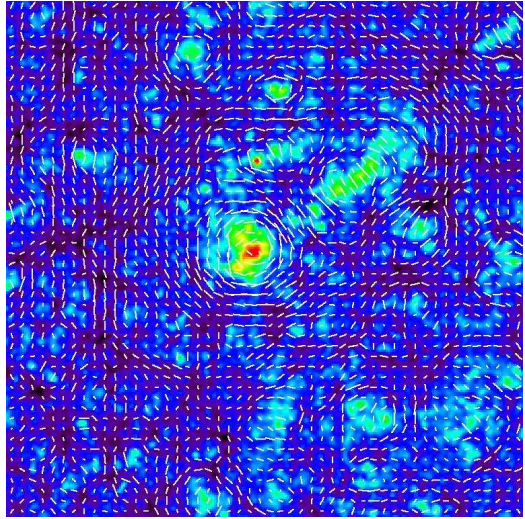,width=2.7in}}
\caption{Cosmic shear field (white ticks) superimposed on the projected mass distribution 
from a cosmological N-body simulation: overdense regions are bright,
underdense regions are dark. Note how the shear field is correlated with the
foreground mass distribution.  Figure courtesy of T. Hamana.  }
\label{fig:Jain_whiskers}
\end{figure}

The gravitational bending of light by structures in the Universe distorts or
shears the images of distant galaxies; see Fig.~\ref{fig:Jain_whiskers}. 
This distortion allows the distribution of dark matter and its evolution 
with time to be measured, thereby probing the influence of dark energy on 
the growth of structure.

The
statistical signal due to gravitational lensing by large-scale structure is
termed ``cosmic shear.''  The cosmic shear field at a point in the sky is
estimated by locally averaging the shapes of large numbers of distant galaxies.
The primary statistical measure of the cosmic shear is the shear angular power
spectrum measured as a function of source-galaxy redshift $z_s$.  (Additional
information is obtained by measuring the correlations between shears at
different redshifts or with foreground lensing galaxies.)  The shear angular
power spectrum is \citep{Kaiser92,hujain03}
\begin{equation}
P^\gamma_{\ell}(z_s) = \int_{0}^{z_s} dz  {H(z) \over d_A^2(z)} |W(z,z_s)|^2
P_\rho\left (k={\ell \over d_A(z)}; z\right )\,,
\label{eqn:Limber}
\end{equation}
where $\ell$ denotes the angular multipole, the weight function $W(z,z_s)$ is the efficiency 
for lensing a population of source galaxies and is determined by the distance distributions 
of the source and lens galaxies, 
and $P_\rho(k,z)$ is the power spectrum of density
perturbations.  

As with clusters, the dark-energy sensitivity of the shear angular power
spectrum comes from two factors: {\it geometry}---the Hubble parameter, the
angular-diameter distance, and the weight functions; and {\it growth of
structure}---through the evolution of the power spectrum of density
perturbations.  It is also possible to separate these effects and extract a
purely geometric probe of dark energy from the redshift dependence of
galaxy-shear correlations \citep{JainTaylor,Bernstein_Jain}.
The three-point correlation of cosmic shear is also sensitive to 
dark energy \citep{Takada_Jain}.

\begin{figure}[!t]
\centerline{\psfig{file=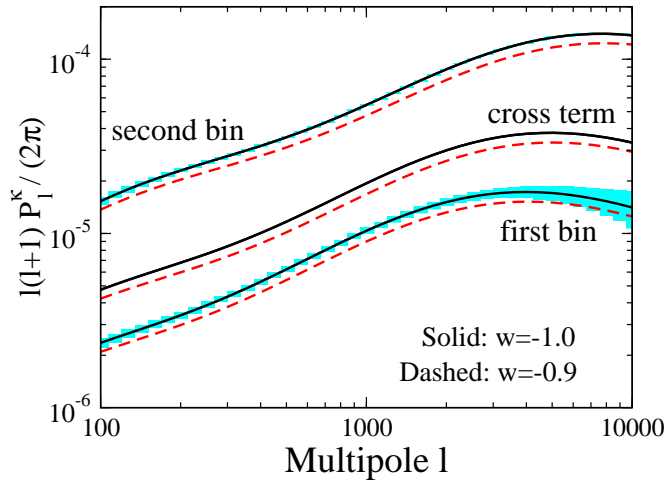,width=3.6in,height=2.8in,angle=-90}}
\caption{
Cosmic shear angular power spectrum and statistical errors 
expected for LSST 
(see \S \ref{projects}) for $w=-1$ and $-0.9$. For 
illustration, 
results are shown for source galaxies in two broad redshift bins,  
$z_s=0-1$ (first bin) and $z_s=1-3$ (second bin); the cross-power 
spectrum between the two bins (cross term) is shown without the 
statistical errors.
}
\label{fig:P_kappa_tomo}
\end{figure}

The statistical uncertainty in measuring the shear power spectrum on
large scales is \citep{Kaiser92}
\begin{equation}
\Delta P^\gamma_\ell = \sqrt{\frac{2}{(2\ell+1)f_{\rm sky}} }
\left[ P^\gamma_\ell +\frac{\sigma^2(\gamma_i)}{n_{\rm eff}} \right]~~,
\label{eqn:power_error}
\end{equation}
where $f_{\rm sky}$ is the fraction of sky area covered by the survey,
$\sigma^2(\gamma_i)$ is the variance in a single component of the
(two-component) shear, and $n_{\rm eff}$ is the effective number density per
steradian of galaxies with well-measured shapes. The first term in brackets,
which dominates on large scales, comes from cosmic variance of the mass
distribution, and the second, shot-noise term results from both the variance
in galaxy ellipticities (``shape noise'') and from shape-measurement errors
due to noise in the images. Fig.~\ref{fig:P_kappa_tomo} shows the dependence
on the dark energy of the shear power spectrum and an indication of the 
statistical errors expected for a survey such as LSST, assuming  
a survey area of 15,000 sq. deg. and effective source galaxy 
density of 
$n_{\rm eff}=30$ galaxies per sq. arcmin.

Systematic errors in weak lensing measurements arise from a number of sources
\citep{Huterer06}: incorrect shear estimates, uncertainties in galaxy
photometric redshift estimates, intrinsic correlations of galaxy shapes, and
theoretical uncertainties in the mass power spectrum on small scales.  The
dominant cause of galaxy shape measurement error in current lensing surveys is
the anisotropy of the image point spread function (PSF) caused by optical and
CCD distortions, tracking errors, wind shake, atmospheric refraction, etc. 
This error can be diagnosed since there are geometric 
constraints on the shear patterns that can be produced by lensing that 
are not respected by systematic effects. 
A second kind of shear measurement error arises from miscalibration of the
relation between measured galaxy shape and inferred shear,
arising from inaccurate correction for the
circular blurring of galaxy images due to atmospheric seeing. 
Photometric redshift errors impact shear power spectrum estimates primarily
through uncertainties in the scatter and bias of photometric redshift estimates
in redshift bins \citep{Huterer06,Ma06}.  
Any tendency of galaxies to align with their neighbors --- or to align with
the local mass distribution --- can be confused with alignments caused by
gravitational lensing, thus biasing dark energy determinations
\citep{Hirata04,Heymans}.  Finally, uncertainties in the theoretical mass power
spectrum on small scales could complicate attempts to use the high-multipole
($\ell \gtrsim$ several hundred) shear power spectrum to constrain dark
energy. Fortunately, weak lensing surveys should be able to internally
constrain the impact of such effects \citep{zentner07}.

\subsection{Other probes}
\label{sec:otherprobes}
While the four methods discussed above have the most probative
power, a number of other methods have been proposed, offering the possibility
of additional consistency checks. The Alcock-Paczynski test exploits the fact
that the apparent shapes of intrinsically spherical cosmic structures depend on 
cosmology \citep{alcock}. Since spatial clustering is
statistically isotropic, the anisotropy of the two-point correlation function
along and transverse to the line of sight has been proposed for this test,
e.g., using the Lyman-alpha forest \citep{hui99}.

Weak lensing of the CMB anisotropy by foreground clusters, in combination 
with lensing of galaxies, provides a potential geometric probe of 
dark energy \citep[e.g.,][]{huholzvale}.

The Integrated Sachs-Wolfe (ISW) effect provided a confirmation of cosmic
acceleration, cf. \S 4.1.2. ISW impacts the large-angle structure of the CMB
anisotropy, but low-$\ell$ multipoles are subject to large cosmic variance,
limiting their power.  Nevertheless, ISW is of interest because it
may be able to show the imprint of large-scale dark-energy perturbations
\citep{Coble,Hu_Scranton}.

Gravitational radiation from inspiraling binary neutron stars or black holes
can serve as ``standard sirens'' to measure absolute distances. If their redshifts 
can be determined, then they
could be used to probe dark energy through the Hubble diagram
\citep{Dalal06}.

Long-duration gamma-ray bursts have been proposed as standardizable candles
\citep[e.g.,][]{schaefer03}, but their utility as cosmological distance
indicators that could be competitive with or complementary to SNe Ia has yet
to be established \citep{friedman05}. The angular size-redshift relation 
for double radio galaxies has also been used to derive cosmological 
constraints that are consistent with dark energy \citep{Guerra}. 
The optical depth for strong
gravitational lensing (multiple imaging) of QSOs or radio sources has been
proposed \citep{fukugita92} and used
\citep[e.g.,][]{mitchell05,chae07} to provide independent evidence for
dark energy, though these measurements depend on modeling the 
density profiles of lens galaxies.

Polarization
measurements from distant galaxy clusters in principle provide a sensitive
probe of the growth function and hence dark energy
\citep{Coo_Hut_Bau}. The relative ages of galaxies at different redshifts, if they can be 
determined reliably, provide a measurement of $dz/dt$ and, from
Eq.~(\ref{eq:cosmic_time}), measure the expansion history directly
\citep{Jimenez_Loeb}. Measurements of the abundance of lensed arcs in
galaxy clusters, if calibrated accurately, provide a probe of dark energy
\citep{Meneghetti_04}.

As we have stressed, there is every reason to expect that at early
times dark energy was but a tiny fraction of the energy density.  Big bang 
nucleosynthesis 
and CMB anisotropy have been used to test this prejudice, and current data
already indicate that dark energy at early times contributes no
more than $\sim 5\%$ of the total energy density
\citep{Bean_Han_Mel,Doran_earlyDE}.

\subsection{Role of the CMB}
While the CMB provides precise cosmological constraints, 
by itself it has little power to probe dark energy (see Fig. \ref{fig:Omw_and_w0wa}).  The reason
is simple: the CMB provides a single snapshot of the Universe at a time when dark
energy contributed but a tiny part of the total energy density (a part in
$10^9$ for vacuum energy).  Nonetheless, the CMB plays a critical supporting role
by determining other cosmological parameters, such as the spatial curvature
and matter density, to high precision, thereby considerably strengthening the
power of the methods discussed above, cf. Fig.~\ref{fig:allen06}. It also 
provides the standard ruler for BAO measurements. Data from the 
Planck CMB mission, scheduled for launch in 2008, will complement those from 
dark energy surveys. If the Hubble parameter can be directly measured to better than 
a few percent, in combination with Planck it would also 
provide powerful dark energy constraints \citep{Hu04book}.

\subsection{Probing new gravitational physics}
\label{sec:DE_vs_MG}

In \S 5.2 we discussed the possibility that cosmic acceleration could be
explained by a modification of General Relativity on large scales. How can
we distinguish this possibility from dark energy within GR and/or test
the consistency of GR to explain cosmic acceleration?  
Since modified gravity can change both the Friedmann equation and the 
evolution of density perturbations, 
a strategy for testing the consistency of GR and dark energy as the 
explanation for acceleration is to compare results from the geometric (expansion history) 
probes, e.g., SNe or BAO, with those 
from the probes sensitive to the growth of structure, e.g., clusters or weak lensing. 
Differences between the two could be evidence for the need to modify GR 
\citep{Knox_Song_Tyson}.
A first application of this idea to current data shows that standard GR passes
a few modest consistency tests \citep{Chu_Knox,Wang_Hui}.

Finally, any modification of gravity may have observable effects beyond cosmology, and 
precision 
solar system tests can provide important additional constraints \citep[e.g.,][]{luess}.

\subsection{Summary and comparison}\label{sec:comparison_methods}

\begin{figure}
\centerline{
\psfig{file=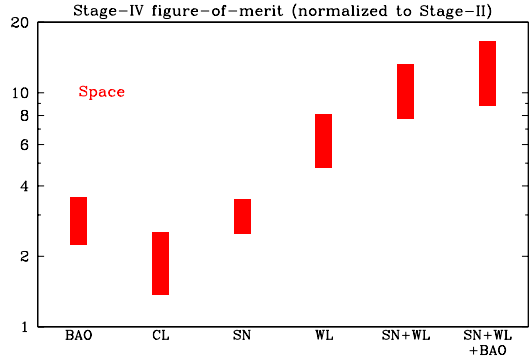,width=3.2in}
}
\caption{Relative statistical power of different dark energy 
space probes, separately 
and in combination, in constraining the DETF Figure of Merit (FoM)
(see \S \ref{appendix}).  Bars indicate 
estimated range of increase (allowing for uncertainties in systematic errors) in 
the FoM relative to present experiments. 
Adopted from the DETF
report \citep{DETF}.  }
\label{fig:DETF_FoM} 
\end{figure}

Four complementary 
cosmological techniques have the power to probe dark energy with 
high precision and thereby advance our understanding of cosmic acceleration:
Weak Gravitational Lensing (WL); type Ia supernovae (SN); Baryon Acoustic
Oscillations (BAO); and Galaxy Clusters (CL).  To date, constraints upon 
the dark energy equation-of-state parameter have come from combining the 
results of two or more techniques, e.g., SN+BAO+CMB (see Fig. \ref{fig:allen06}) 
or BAO+CMB (see Table \ref{tab:cosmo_parameters}), in order 
to break cosmological parameter degeneracies. In the future, each of these 
methods, in combination with CMB information that constrains 
other cosmological parameters, will provide powerful 
individual constraints on dark energy; collectively, they should 
be able to approach percent-level precision on $w$ at its best-constrained 
redshift, i.e., $w_p$ (see Fig. \ref{fig:Omw_and_w0wa}). 

Table \ref{tab:probes} summarizes these four dark energy probes,
their strengths and weaknesses and primary systematic
errors. Fig.~\ref{fig:DETF_FoM} gives a visual impression of the statistical
power of each of these techniques in constraining dark energy, showing how
much each of them could be expected to improve our present knowledge of $w_0$
and $w_a$ in a dedicated space mission \citep{DETF}.

\begin{table}[!h]
  \begin{minipage}{5in}
    \caption{Comparison of dark energy probes.}\vspace{0.2cm}
    \label{tab:probes}
    \begin{tabular}{llll} \hline\hline
\rule[-3mm]{0mm}{8mm}Method & Strengths & Weaknesses & Systematics \\ \hline
%%%%%%%%%%%%%%%%%%%%%%%%%%%%%
% enter the 1st row, 1st line
%%%%%%%%%%%%%%%%%%%%%%%%%%%%%
WL & growth+geometric, & CDM assumption & image quality,                              \\[-0.5ex]
% 2nd line of 1st row
& statistical power & & photo-z               \\\hline
%%%%%%%%%%%%%%%%%%%%%%%%%%%%%
%%%%%%%%%%%%%%%%%%%%%%%%%%%%%
SN & purely geometric, & standard candle & evolution,                                       \\[-0.5ex]
% 2nd line of 4th  row
& mature & assumption & dust              \\\hline
%%%%%%%%%%%%%%%%%%%%%%%%%%%%%
BAO & largely geometric, & large samples & bias,   \\[-0.5ex]
& low systematics & required & non-linearity                           \\\hline
%%%%%%%%%%%%%%%%%%%%%%%%%%%%%
CL & growth+geometric,  & CDM assumption & determining mass,                                \\[-0.5ex]
% 2nd line of 3rd row
& X-ray+SZ+optical &  & selection function   \\\hline
%%%%%%%%%%%%%%%%%%%%%%%%%%%%%
\hline
    \end{tabular}
  \end{minipage}
\end{table}

\section{DARK ENERGY PROJECTS}
\label{projects}

A diverse and ambitious set of projects to probe dark energy are in
progress or being planned. Here we provide a brief overview 
of the observational landscape.
With the exception of experiments at the LHC
that might shed light on dark energy through discoveries about supersymmetry
or dark matter, all planned experiments involve cosmological observations. 
Table \ref{tab:surveys}  
provides a representative sampling, not a comprehensive listing, of projects that are 
currently proposed or under construction and does not 
include experiments that have already 
reported results. All of these projects share the common feature of 
surveying wide
areas to collect large samples of objects --- galaxies, clusters, or
supernovae. 

The Dark Energy Task Force (DETF) report \citep{DETF} classified dark energy
surveys into an approximate sequence: on-going projects, either taking data or
soon to be taking data, are Stage II; near-future, intermediate-scale projects
are Stage III; and larger-scale, longer-term future projects are designated
Stage IV. More advanced stages are in general expected to deliver tighter dark
energy constraints, which the DETF quantified using the $w_0$-$w_a$ figure of
merit (FoM) discussed in the Appendix (\S \ref{sec:FoM}). Stage III experiments
are expected to deliver a factor $\sim 3-5$ improvement in the DETF FoM
compared to the combined Stage II results, while Stage IV experiments should
improve the FoM by roughly a factor of 10 compared to Stage II, though these
estimates are only indicative and are subject to considerable uncertainties in
systematic errors (see Fig. \ref{fig:DETF_FoM}).

We divide our discussion into ground- and space-based surveys.  Ground-based
projects are typically less expensive than their space-based counterparts and
can employ larger-aperture telescopes.  The discovery of dark energy and many of the
subsequent observations to date have been dominated by ground-based
telescopes. On the other hand, HST (high-redshift SN observations), Chandra
(X-ray clusters), and WMAP CMB observations have played critical roles in
probing dark energy. While more challenging to execute, space-based surveys
offer the advantages of observations unhindered by weather and by the
scattering, absorption, and emission by the atmosphere, stable observing
platforms free of time-changing gravitational loading, and the ability to
continuously observe away from the sun and moon. They therefore have the
potential for much improved control of systematic errors.

\subsection{Ground-based surveys}

A number of projects are underway to detect clusters and probe dark energy using the
SZE (see Sec.~\ref{sec:clusters}). These surveys are
coordinated with optical surveys that can determine cluster redshifts. The
Atacama Pathfinder EXperiment (APEX) survey in Chile 
will cover up to 1000 square degrees. The largest of these projects
are the Atacama Cosmology Telescope (ACT) and the South Pole Telescope
(SPT), the latter of which will carry out a 4,000 square degree survey.

A number of optical imaging surveys are planned or proposed which can study
dark energy through weak lensing, clusters, and angular BAO using a single
wide-area survey.  These projects use telescopes of intermediate to large
aperture and wide field-of-view, gigapixel-scale CCD cameras, and are deployed
at the best astronomical sites in order to obtain deep galaxy photometry and
shape measurements. They deliver photometric-redshift information through color
measurements using multiple passbands. The ESO VLT Survey Telescope (VST) on
Cerro Paranal will carry out public surveys, including the 1500 sq. deg. KIDS
survey and a shallower, 5000 sq. deg. survey (ATLAS). The Panoramic Survey
Telescope and Rapid Response System (Pan-STARRS)-1 uses a 1.8-m wide-field
telescope to carry out several wide-area surveys from Haleakala; in the future,
they hope to deploy $4 \times 1.8$-m telescopes at Mauna Kea in
Pan-STARRS-4. The Dark Energy Survey (DES) will use a new 3 sq. deg.  imager
with red-sensitive CCDs on a 4-m telescope 
at Cerro Tololo Inter-American Observatory (CTIO) in Chile to
carry out a 5,000 sq. deg. survey in 5 optical passbands, covering the same
survey area as the SPT and partnering with the ESO VISTA Hemisphere Survey
which will survey the same area in 3 near-infrared bands.  Hyper Suprime-Cam is
a new wide-field imager planned for the Subaru telescope on Mauna Kea that will
be used to carry out a deep survey over 2000 sq. deg.  The Advanced
Liquid-mirror Probe of Asteroids, Cosmology and Astrophysics (ALPACA) is a
proposed rotating liquid mercury telescope that would repeatedly survey a long,
narrow strip of the sky at CTIO.  The most ambitious of these projects is the
Large Synoptic Survey Telescope (LSST), which would deploy a multi-Gigapixel
camera with 10 sq. deg.  field-of-view on a new telescope on Cerro Pachon in
Chile to survey 15,000 sq.  deg. over 10 years.

\begin{table*}
  \begin{minipage}{5in}
    \caption{Dark energy projects proposed or under construction. Stage refers 
to the DETF time-scale classification.\vspace{0.2cm}}
    \label{tab:surveys}
    \begin{tabular}{llll} \hline\hline
\rule[-3mm]{0mm}{8mm} Survey & Description & Probes & Stage \\ \hline
%%%%%%%%%%%%%%%%%%%% GROUND %%%%%%%%%%%%%%%%%%%%%%%
Ground-based: && \\
ACT        & SZE, 6-m                                  & CL & II  \\
APEX       & SZE, 12-m                                 & CL & II  \\
SPT        & SZE, 10-m                                 & CL & II  \\
VST        & Optical imaging, 2.6-m                   & BAO,CL,WL & II \\
Pan-STARRS 1(4) & Optical imaging, 1.8-m($\times 4$)  & All & II(III) \\
DES        & Optical imaging, 4-m                     & All & III \\
Hyper Suprime-Cam & Optical imaging, 8-m              & WL,CL,BAO & III \\
ALPACA     & Optical imaging, 8-m                     & SN, BAO, CL & III \\
LSST       & Optical imaging, 6.8-m                   & All & IV \\
AAT WiggleZ& Spectroscopy, 4-m                        & BAO & II \\
HETDEX     & Spectroscopy, 9.2-m                      & BAO & III \\
PAU        & Multi-filter imaging, 2-3-m                     & BAO & III \\
SDSS BOSS  & Spectroscopy, 2.5-m                      & BAO & III \\
WFMOS      & Spectroscopy, 8-m                        & BAO & III \\
HSHS       & 21-cm radio telescope                    & BAO & III \\
SKA        & km$^2$ radio telescope                   & BAO, WL & IV \\
\hline
%%%%%%%%%%%%%%%%%%%%  SPACE %%%%%%%%%%%%%%%%%%%%%%%%
Space-based: && \\
{\em JDEM Candidates}                               &&   \\
\ \ ADEPT      & Spectroscopy            & BAO, SN & IV  \\         
\ \ DESTINY    & Grism spectrophotometry & SN      & IV  \\ 
\ \ SNAP       & Optical+NIR+spectro     & All  & IV  \\ 
{\em Proposed ESA Missions}                            &&\\
\ \ DUNE       & Optical imaging         & WL, BAO, CL      &     \\
\ \ SPACE      & Spectroscopy            & BAO     &     \\
\ \ eROSITA    & X-ray                   & CL      &     \\
{\em CMB Space Probe}                                 && \\
\ \ Planck     & SZE                      & CL      &     \\
{\em Beyond Einstein Probe}                           && \\
\ \ Constellation-X & X-ray              & CL      & IV  \\
\hline\hline
    \end{tabular}
  \end{minipage}
\end{table*}

Several large spectroscopic surveys have been designed to detect baryon
acoustic oscillations by measuring $\sim 10^5-10^9$ galaxy and QSO redshifts
using large multi-fiber spectrographs.  WiggleZ is using the Anglo-Australian
Telescope to collect spectra of 400,000 galaxies in the redshift range
$0.5<z<1$.  The Baryon Oscillation Sky Survey (BOSS) proposes to use the SDSS
telescope in New Mexico to measure galaxy spectra out to $z=0.6$.  The Hobby
Eberly Telescope Dark energy EXperiment (HETDEX) plans to target Ly-{$\alpha$}
emitters at higher redshift, $2\lesssim z\lesssim 4$. The Wide-Field
Multi-Object Spectrograph (WFMOS), proposed for the Subaru telescope, would
target galaxies at $z\lesssim 1.3$ and Lyman-break galaxies at $2.5\lesssim
z\lesssim 3.5$. The Physics of the Accelerating Universe (PAU) is a Spanish 
project to deploy a wide-field camera with a large number of narrow filters 
to measure coarse-grained galaxy
spectra out to $z=0.9$.

Finally, the proposed Square Kilometer Array (SKA), an array of radio antennas
with unprecedented collecting area, would probe dark energy using baryon acoustic 
oscillations and weak lensing of galaxies via measurements of the 21-cm line
signature of neutral hydrogen (HI). The 
Hubble Sphere Hydrogen Survey (HSHS) aims to carry out a 21-cm BAO 
survey on a shorter timescale.

\begin{figure}[!ht]
\centerline{\psfig{file=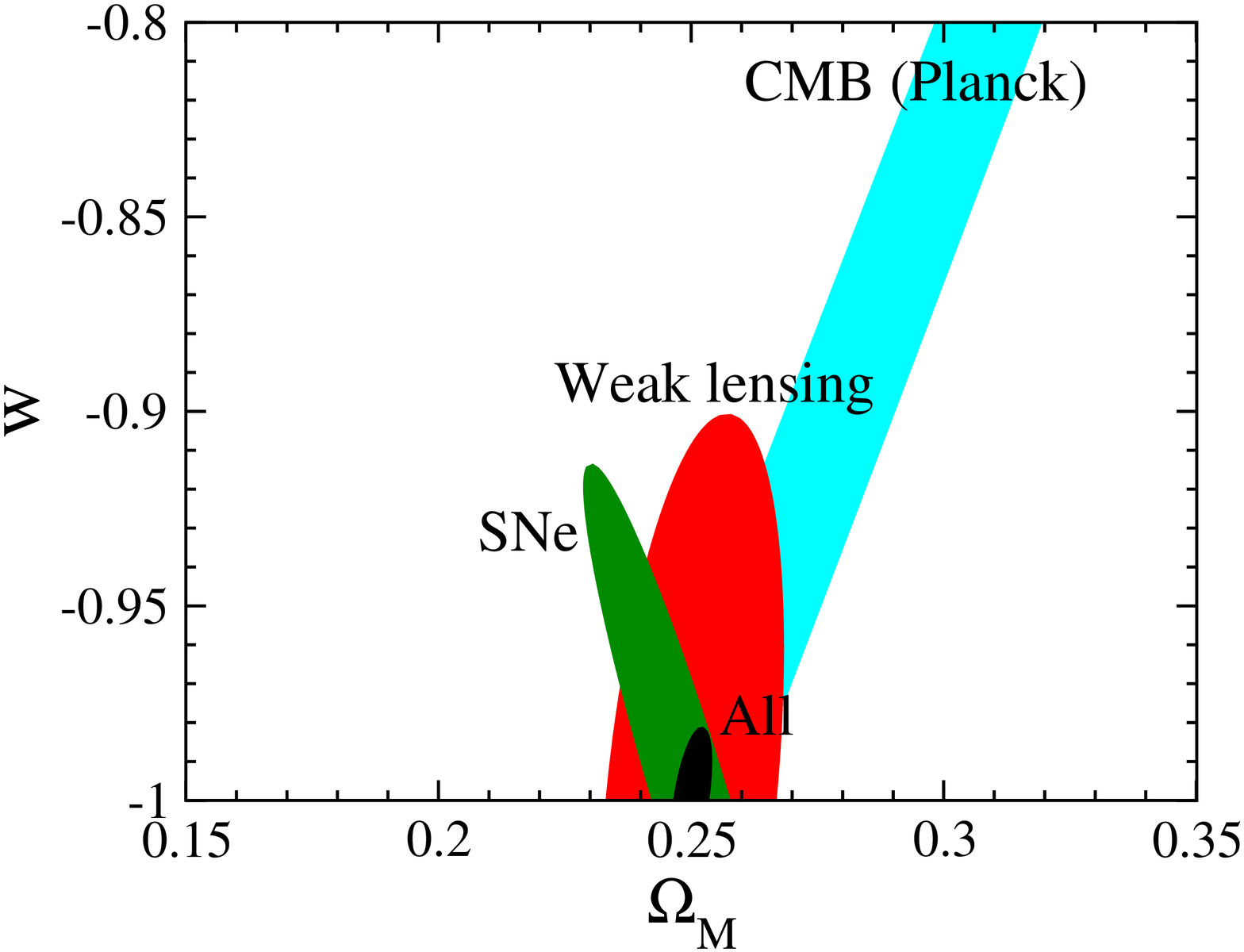,height=2.3in,angle=-90}
\hspace{-0.4cm}
\vspace{-0.2cm}
\psfig{file=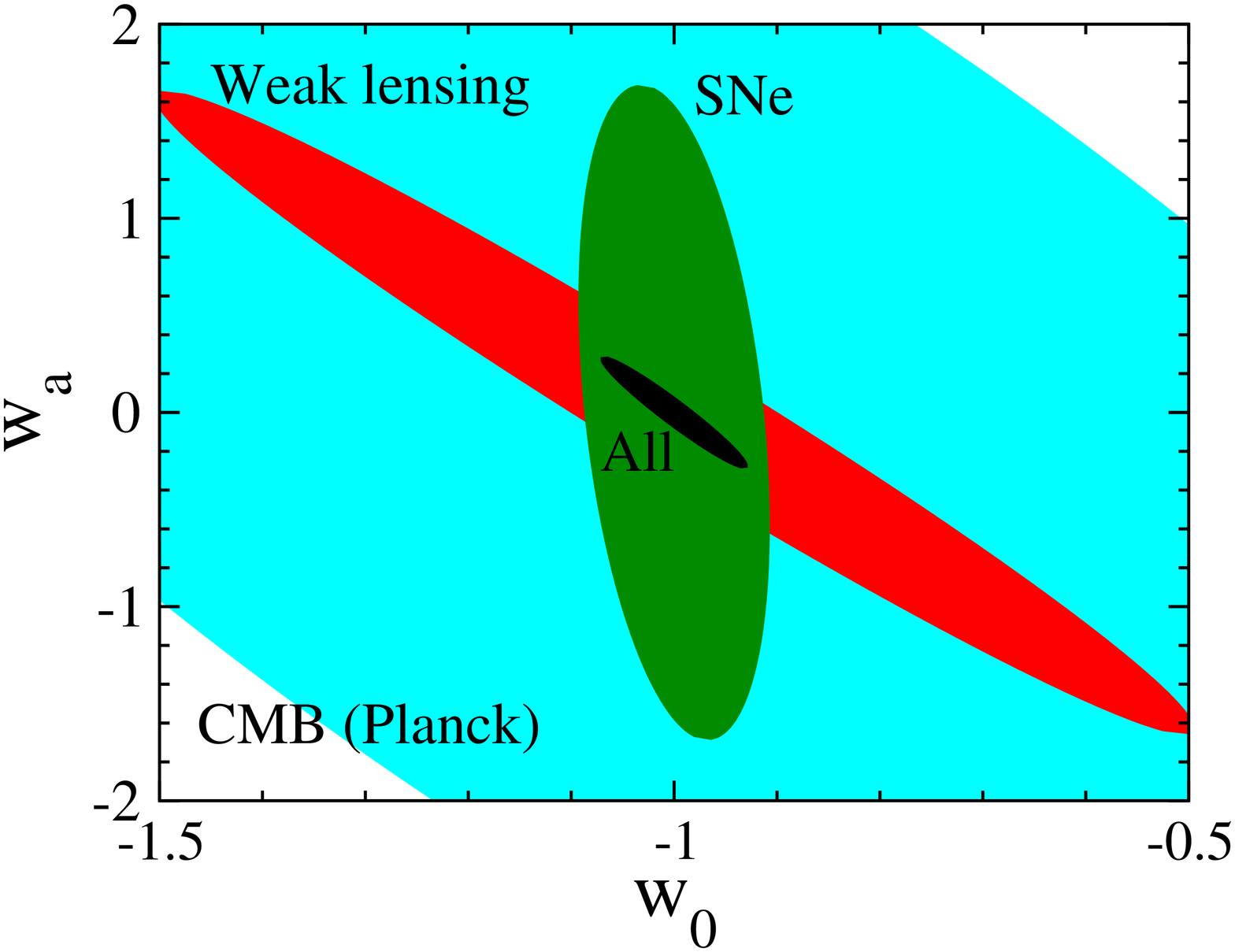,height=2.3in,angle=-90}
}
\caption{Illustration of forecast constraints on dark energy parameters. 
Shown are 68\% C.L.\ uncertainties for one version of the proposed SNAP
experiment, which combines a narrow-area survey of 2000 SNe to $z=1.7$ and a
weak lensing survey of 1000 sq. deg.  {\it Left panel:} Constraints in the
$\Omega_{\rm M}$-$w$ plane, assuming constant $w$; the vertical axis can also
be interpreted as the pivot value $w_p$ for a time-varying equation of
state. {\it Right panel:} Constraints in the $w_0$-$w_a$ plane for time-varying
dark energy equation of state, marginalized over $\Omega_{\rm M}$ for a flat
Universe.  }
\label{fig:Omw_and_w0wa}
\end{figure}

\subsection{Space-based surveys}

Three of the proposed space projects are candidates for the Joint Dark Energy
Mission (JDEM), a joint mission of the U.S. Department of Energy (DOE) and the
NASA Beyond Einstein program, targeted at dark energy science.
SuperNova/Acceleration Probe (SNAP) proposes to study dark energy using a
dedicated 2-m class telescope. With imaging in 9 optical and near-infrared passbands
and follow-up spectroscopy of supernovae, it is principally designed to probe
SNe Ia and weak lensing, taking advantage of the excellent optical image
quality and near-infrared transparency of a space-based platform.
Fig.~\ref{fig:Omw_and_w0wa} gives an illustration of the statistical
constraints that the proposed SNAP mission could achieve, by combining SN and
weak lensing observations with results from the Planck CMB mission.  This
forecast makes use of the Fisher information matrix described in the Appendix
(\S \ref{sec:Fisher}).  The Dark Energy Space Telescope (DESTINY) would use a
similar-size telescope with a near-infrared grism spectrograph to study
supernovae.  The Advanced Dark Energy Physics Telescope (ADEPT) is a
spectroscopic mission with the primary goal of constraining dark energy via
baryon acoustic oscillations at $z\sim 2$ as well as supernovae. Another
proposed mission within the NASA Beyond Einstein program is Constellation-X,
which could observe X-ray clusters with unprecedented sensitivity.

There is one European Space Agency (ESA) mission nearing launch and two
concepts under study. The Planck mission, planned for launch in late 2008, 
in addition to pinning down other cosmological parameters important for 
dark energy, will detect thousands of
galaxy clusters using the SZE.  Dark Universe Explorer
(DUNE) and SPACE are optical missions to study dark energy using weak lensing
and baryon acoustic oscillations, respectively. Finally, the extended ROentgen
Survey with an Imaging Telescope Array (eROSITA), a German-Russian
collaboration, is a planned X-ray telescope that will study dark energy using
the abundance of X-ray clusters.

\section{DARK ENERGY \& COSMIC DESTINY}
\label{destiny}

\begin{figure}[!ht]
\centerline{\psfig{file=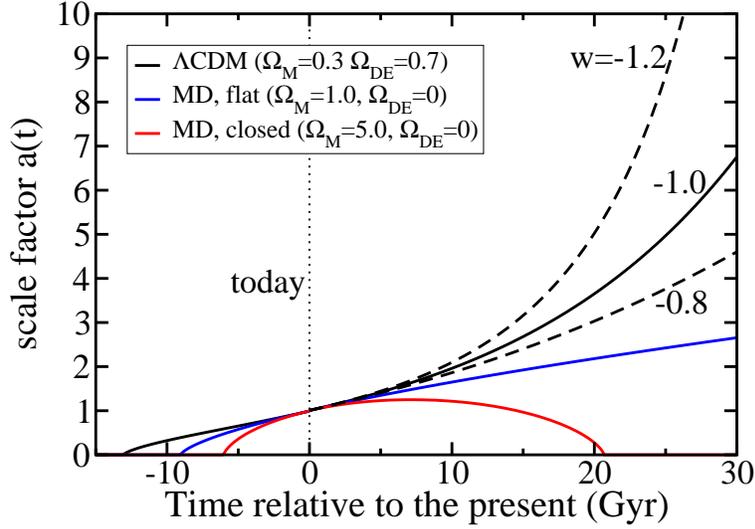,height=3.3in,angle=-90}}
\caption{Evolution of the FRW scale factor in models with and without dark energy. Upper four curves are for flat models. Dashed curves denote models 
with $w=-0.8$ or $-1.2$ and $\Omega_{\rm M}=0.3$. 
MD denotes matter-dominated models.}
\label{fig:fate}
\end{figure}

One of the first things one learns in cosmology is that geometry is destiny: a closed (positively curved) Universe eventually recollapses, and an open (flat or negatively curved) Universe expands forever.  Provided the Universe contains only matter and $\Lambda= 0$, this follows directly from Eq.~(\ref{eq:feq}).  The presence of dark energy severs this well known connection between geometry and destiny and raises fundamental issues involving the distant future of our Universe \citep{Krauss_Turner}.

To illustrate the geometry-destiny connection, we can 
rewrite Eq.~(\ref{eq:feq}) in terms of an effective potential and a kinetic-energy term, 
\begin{equation}
V_{\rm eff} (a) + {\dot a}^2 = 0 \qquad ~~ V_{\rm eff} (a) =  k - \Omega_0H_0^2a^{-(1+3w_{\rm T})}~,
\end{equation}
where $w_{\rm T}$ is the ratio of the total pressure to the  
total energy density (including all components). If $w_{\rm T}>-1/3$, as would be the case with only matter and radiation, then the second term in $V_{\rm eff}$ increases monotonically from $-\infty$ to $0$ as $a$ goes from 0 to $\infty$, which means that $V_{\rm eff}$ rises from $-\infty$ to $k$.  For $k>0$, there is a value of $a$ where $V_{\rm eff}\rightarrow 0$, at which point $\dot a$ must go to zero and $a$ achieves its maximum value. For $k=0$, $\dot a$ only vanishes for $a \rightarrow \infty$; and for $k<0$, $\dot a$ remains positive even as $a \rightarrow \infty$.

With dark energy there is a new twist:  since the dark energy density decreases more slowly than that of matter or radiation, as the Universe expands dark 
energy eventually dominates 
the second term in $V_{\rm eff}$.  Thereafter, $V_{\rm eff}$ decreases monotonically, since $w_{\rm T} \simeq w_{\rm DE}<-1/3$, approaching $-\infty$ as $a \rightarrow \infty$.  Provided that $\rho_{\rm DE}>0$ and that $w_{\rm DE}$ 
remains negative, if the scale factor becomes large enough for dark energy to dominate, which happens unless $\Omega_{\rm M} >1 \gg \ode$, then the Universe will expand forever, irrespective of $k$. 

If dark energy is vacuum energy, acceleration will continue, and
the expansion will become exponential, leading to a ``red out'' of the Universe. 
To see this, consider the comoving distance to fixed redshift $z$ at time $t$ 
during the epoch of exponential expansion:
\begin{equation}
r(z,t)=\int_{a(t)/(1+z)}^{a(t)}\,{da \over a^2H} \simeq {zH_0^{-1}\exp [-H_0(t-t_0)]} ~.
\label{eq:redout}
\end{equation}
The exponential decrease of this distance implies 
that the number of galaxies below a fixed 
redshift shrinks exponentially. 
By contrast, in the Einstein-de Sitter model with 
$\Omega_{\rm M}=1$, this distance increases as $t^{1/3}$, so that the number of 
galaxies  with redshift less than a fixed value grows slowly. 
Alternatively, Eq. \ref{eq:redout} implies that the redshift for a galaxy at current  
distance $r$ grows exponentially.  
Galaxies beyond the Local Group, $r \gtrsim 1-2$ Mpc,  
will be redshifted beyond detectability 
on a timescale of $t-t_0 \sim 100$ Gyr \citep[e.g.,][]{busha}. The Milky Way 
will remain gravitationally bound to the Local Group, which will appear
as a static, ``island Universe.'' Even the CMB, the other key evidence of 
a once-hot, expanding Universe, will be redshifted to undetectability \citep{Krauss_Scherrer}.

If dark energy is a scalar field, then eventually the field relaxes to the minimum of its potential; see Fig~\ref{fig:Vphi}.  If the minimum of the potential energy is precisely zero, the Universe will again become matter dominated and return to decelerated expansion, restoring the link between geometry and destiny.  If the minimum of the scalar field potential has negative energy density, the energy of dark matter and of scalar field energy will eventually cancel, leading to recollapse, irrespective of $k$.  If the potential energy at the minimum is positive and larger than a critical value that depends on $\Omega_{\rm M}$ (the critical value is zero for $\Omega_{\rm M} \leq 1$ and small for $\Omega_{\rm M}>1$), then 
accelerated expansion will eventually ensue again and as discussed above, the Universe will experience a ``red-out.''  These possibilities are illustrated in Fig.~\ref{fig:fate}.

Finally, the possibility of $w_{\rm DE}<-1$ deserves special comment.  In this case,
the energy density of dark energy actually increases with time, $\rho_{DE}
\propto a^\beta$, where $\beta \equiv -3(1+w) > 0$.  In turn, the scale factor
grows very rapidly and reaches infinite size in a finite time:
\begin{equation}
a(t) \simeq {1\over \left[ 1 - \beta H_0(t-t_0)/2 \right]^{2/\beta}}\,,\qquad
(t_\infty - t_0) \simeq {2\over \beta H_0}\,.
\end{equation}
This is the ``big rip'' scenario of \citet{bigrip}.  
Once the density of dark energy exceeds that of any structure -- from
clusters to atoms -- that structure is torn apart.

The presence of dark energy severs the simple relation between geometry and destiny, links destiny to an understanding of dark energy, raises the specter of a bleak future for cosmologists, and raises a deep question \citep{Krauss_Turner}: can we ever determine the future of the Universe with certainty? As a thought experiment, ignore the current epoch of accelerated expansion and imagine instead that the Universe has been determined to be matter dominated and that it is flat. We might be tempted to conclude that the Universe will expand forever at an ever-decreasing rate.  However, no matter how precise our measurements are, there could be a small cosmological constant lurking below the threshold of detectability.  For example, if the 
vacuum energy density were one billionth of the present matter density, after a factor of 1000 in expansion vacuum energy would come to dominate.  If it were positive, exponential expansion would eventually ensue; if negative, the Universe would ultimately recollapse.  Only a fundamental understanding of the constituents of the Universe and their relative abundances could deliver certainty about the destiny of the Universe.

\section{CONCLUDING REMARKS}
\label{conclusion}
Ten years after its discovery, the acceleration of the expansion 
of the Universe is now firmly established. The physical origin of this phenomenon, however, 
remains a deep mystery, linked to other important problems in physics and astronomy.  
At present, the simplest explanation, vacuum energy, is consistent 
with all extant data, but theory provides no understanding 
of why it should have the requisite small value. Probing the history 
of cosmic expansion with much greater precision (few percent vs. current 10\%) offers the best hope of pointing us down the correct path to a solution. An 
impressive array of experiments with that aim are underway or planned, 
and we believe that significant progress will be made in the next fifteen years.

We conclude with our list of the ten important take-home facts about cosmic acceleration and dark energy, followed
by our views on the key open issues and challenges for the future.

\subsection{Take-home facts}

\subsubsection{Strong evidence for accelerated expansion.} 
Since the SN discovery of acceleration, 
several hundred supernovae have been observed over a broader range of 
redshifts, substantially strengthening the case both statistically and 
by reducing sources of systematic error.  Further, independent of GR 
and based solely upon the SN Hubble diagram, there is very strong ($5\sigma$) 
evidence that the expansion of the Universe accelerated recently 
\citep{Shapiro_Turner}.

\subsubsection{Dark energy as the cause of cosmic acceleration.}  
Within GR, accelerated expansion cannot be explained by any known form of matter or energy but can be
accommodated by a nearly smooth form of
energy with large negative pressure, known as dark energy, that accounts for about 75\% of the Universe.

\subsubsection{Independent evidence for dark energy.}  
In the context of the cold dark matter model of structure formation, CMB and
large-scale structure data provide independent evidence that the Universe
contains a smooth form of energy which accounts for about 75\% of 
the total and which only came to dominate after
essentially all of the observed structure had formed.  
Thus, structure formation independently points to a negative-pressure (with $w \lesssim -1/3$), dark energy accounting for the bulk of the Universe. 

\subsubsection{Vacuum energy as dark energy.}  
The simplest explanation for dark energy is the energy associated with the vacuum;   
it is mathematically equivalent to a cosmological constant.  However, 
all attempts to compute the vacuum energy density from 
the zero-point energies of all quantum fields yield a result that is 
many orders of magnitude too large or infinite.

\subsubsection{Current observational status.}  
Taken together, all the current data provide strong evidence for the existence of dark energy; 
they constrain the fraction of critical density contributed by dark energy, $0.76\pm 0.02$,
and the equation-of-state parameter, $w\approx -1 \pm 0.1$ (stat) $\pm
0.1$ (sys), assuming that $w$ is constant. This implies that the Universe
began accelerating at redshift $z \sim 0.4$ and age $t \sim 10$ Gyr. These 
results are robust -- data from any one method can be removed 
without compromising the constraints -- and they are not substantially
weakened by dropping the assumption of spatial flatness.  Relaxing the assumption that $w$ is constant and parametrizing its variation
as $w(z) = w_0 + w_a (1-a)$, the current observational constraints are considerably weaker, $\ode\approx 0.7\pm 0.15$, $w_0 \approx -1\pm 0.2$, $w_a \approx 0\pm 1$, and provide no evidence for variation of $w$.

\subsubsection{Dark theory:  dark energy or new gravitational physics?}  
There is no compelling model for dark energy.  Beyond vacuum energy,
there are many intriguing ideas, including a new light scalar field 
and the influence of additional spatial dimensions. In many of these 
models, time-varying dark energy is expected.  On the other hand, cosmic acceleration could be a manifestation of gravitational physics beyond GR rather than dark energy.  
While interesting, there is as yet no self-consistent
model for the new gravitational physics that is also consistent with the
large body of data that constrains theories of gravity.

\subsubsection{Dark Destiny.}  
The destiny of the Universe depends crucially upon the
nature of the dark energy.  All three fates -- recollapse or continued
expansion with and without slowing -- are possible.  The existence of dark energy raises
the issue of cosmic uncertainty: can we determine the mass/energy content with
sufficient precision to rule out the possibility that a tiny dark energy component today may dominate in the distant future?

\subsubsection{At the nexus of many mysteries.}  
Because of its multiple close connections to important problems in both physics and astronomy, cosmic acceleration may be the most profound mystery in science.  Its solution could shed light on or be central to unraveling other important puzzles, including the cause of cosmic
inflation, the vacuum-energy problem, supersymmetry and superstrings, neutrino mass, new gravitational physics, and even dark matter.  

\subsubsection{The two big questions.} 
Today, the two most pressing questions about cosmic
acceleration are: Is dark energy something other than vacuum energy?  Does
GR self-consistently describe cosmic acceleration?
Establishing that $w \not= -1$ or that it varies with time, or 
that dark energy clusters,
would rule out vacuum energy.  Establishing that the values of $w$ determined
by the geometric and growth of structure methods are
not equal could point toward a modification of gravity as the cause of
accelerated expansion.

\subsubsection{Probing Dark Energy.}  
An impressive array of space- and
ground-based observations, using SNe, weak lensing, clusters, and baryon
acoustic oscillations, are in progress or are being planned.
They should determine $w_p$, the equation-of-state parameter at the 
redshift where it is best determined,  at the percent level and its
time variation $w_a$ at the 10\% level, dramatically improving our ability to
discriminate between vacuum energy and something more exotic as well as testing
the self-consistency of GR to explain cosmic acceleration.
Laboratory and accelerator-based experiments could also shed light on dark
energy.

\subsection{Open issues and challenges}
\subsubsection{Clustering of dark energy.}

While vacuum energy is uniform, dynamical forms of dark energy can 
be inhomogeneous, making dark energy clustering a potential additional probe of
dark energy.  
However, since dark energy is likely to cluster only weakly and on the largest 
scales, the prospects for clustering as a probe of dark energy are not high.
Nonetheless, discovering that dark energy clusters would rule out vacuum energy.  Current constraints on the clustering of dark energy are weak, and there may be better ideas about measuring dark-energy clustering.

\subsubsection{Dark energy and matter.}
In scalar field models of dark energy, there is a new, 
very light ($m \lesssim H_0 \sim 10^{-33}\,$eV) scalar particle which can couple to matter and thereby give rise to new long-range forces with potentially observable consequences.  
Such an interaction could perhaps help explain the near-coincidence between the present densities of dark matter and dark energy or change the dynamics of dark matter particles, though it is constrained by astrophysical and cosmological observations 
to be of at most gravitational strength \citep{Gradwohl_Frieman,Carroll_quint}. A
coupling to ordinary matter would have even larger observable effects and is highly constrained.

\subsubsection{Describing cosmic acceleration and dark energy.}
In the absence of theoretical guidance,
the equation-of-state parameter $w\equiv p/\rho$ is a convenient way of characterizing
dark energy and its effects on the expansion.
One can instead take a more agnostic approach and
interpret results in terms of the kinematics of the expansion or the energy
density.  Further, it is worth exploring improved
descriptions of dark energy that both yield physical insight and are better matched to the
observations.

\subsubsection{Systematic errors.}
All of the techniques used to probe dark energy are limited by systematic
errors.  The sources of systematic error include:
luminosity evolution and dust extinction uncertainties (for SNe Ia); shape
measurement systematics, photometric redshift errors, and theoretical modeling
of the matter power spectrum (for weak lensing); galaxy biasing, non-linearity, 
and redshift distortions (for BAO); and the uncertain relations between cluster
mass and its observable proxies (for galaxy clusters).  Improvements
in all of these will be critical to realizing the full potential of planned
observations to probe dark energy and will have beneficial effects more broadly in astronomy.

\subsubsection{Dark Energy Theory.}
The grandest challenge of all is a deeper understanding of the cause of cosmic acceleration.  What is called for is not the invention of ad hoc models based upon clever ideas or new potentials, but rather a small number of theoretical models that are well motivated by fundamental physics and that make specific enough predictions to be falsified.

\subsubsection{How much is enough?} Given its profound implications 
and the absence of a compelling theory, 
dark energy is the exemplar of high-risk, high-gain science. Carrying out the 
most ambitious proposed dark energy projects -- JDEM and LSST -- 
to attain percent level precision 
will cost more than one billion dollars. 
While they will yield much tighter parameter constraints, 
there is no guarantee that they will deliver deeper 
understanding of dark energy.  If they are able to exclude vacuum 
energy or demonstrate the inconsistency of GR, the implications would 
be revolutionary. On the other hand, if they yield  
results consistent with vacuum energy, it would constitute an important test of the ``null hypothesis'' and provide a set of cosmological parameters that 
will satisfy the needs for astrophysical cosmology for the foreseeable 
future. In this case, unless there are new theoretical developments 
pointing to different or more decisive probes of compelling dark energy 
theories, there is likely to be  
little enthusiasm for continuing on to even more expensive dark energy projects. 

There is no doubt that pursuing the origin of cosmic acceleration will
continue to be a great intellectual adventure for the next 
fifteen years. Even if these ambitious projects do not solve this riddle, 
they will at least sharpen the problem and 
will certainly produce a wealth of survey data that will benefit many 
areas of astronomy for decades to come.

%\section{ACKNOWLEDGEMENTS}
%\label{acknowledge}

\newpage

\noindent{\bf \large Acknowledgements}

We thank Andy Albrecht, Gary Bernstein, Roger Blandford, Ed Copeland, Wendy Freedman, Don Goldsmith, Wayne Hu, Rocky Kolb, Eric Linder, Adam Riess, Martin White, and Bruce Winstein for helpful comments on an earlier draft of this article. We also acknowledge the Aspen Center for Physics, where part of this review was written. This work was supported in part by the DOE at Fermilab and at 
the University of Chicago and by the KICP NSF Physics Frontier Center 
grant PHY-0114422.

\section{APPENDIX}
\label{appendix}
\subsection{Figure(s) of merit}\label{sec:FoM}

How do we compare the dark energy ``reach'' of different methods and 
different experiments? We cannot quantify the 
probative power of dark energy methods in a strictly model-independent way, 
since we do not know which aspects of the expansion history are 
most important to measure. Nevertheless, some useful
figures of merit (FoMs) have been proposed to facilitate comparison of 
methods and experimental designs. Examples include 
the volume of the uncertainty ellipsoid for the dark
energy parameters or the thickness of the ellipsoid in its narrowest
direction \citep{Hut_Tur_00}. 
In the Fisher matrix approach (\S \ref{sec:Fisher}),
these correspond to the inverse square root of the determinant and the largest
eigenvalue of the Fisher matrix, respectively. A special case of the volume FoM
is the inverse area of the Fisher-matrix-projected ellipse in the $w_0$-$w_a$
plane,
\begin{equation}
{\rm FoM}  \propto [\sigma(w_0)\sigma(w_a)]^{-1} \propto 
\left (\det{F^{w_0 w_a}}\right )^{1/2}~,
\label{eq:FoM}
\end{equation}
where $F^{w_0 w_a}$ is the Fisher matrix projected onto the $w_0$-$w_a$ plane. 
This choice was adopted by
the Dark Energy Task Force as a metric for comparing methods and surveys and is
shown in relative terms for Stage IV space-based experiments in
Fig.~\ref{fig:DETF_FoM}.
The DETF FoM provides a simple yet useful metric for comparison, as it takes
into account the power of experiments to measure the temporal variation of 
$w$. 
For generalizations, see
\citet{Albrecht_Bernstein}.

\subsection{Fisher information matrix}\label{sec:Fisher}

The Fisher information matrix formalism allows a quick and easy way to estimate
errors on cosmological parameters, given errors in observable quantities, and
is particularly useful in experimental design. The Fisher matrix is
defined as the (negative) Hessian of the log-likelihood function $\mathcal{L}$,
\begin{equation}
F_{ij} \equiv 
\left \langle -{\partial^2 \ln\mathcal{L}\over \partial p_i \partial p_j}\right\rangle 
= {\bf \mu}_{,i}^T {\bf C}^{-1}{\bf \mu}_{,j} + {1\over 2}{\rm Tr}
\left [{\bf C}^{-1}{\bf C}_{,i}{\bf C}^{-1}{\bf C}_{,j}\right ].
\end{equation}
The second equality follows by assuming that $\mathcal{L}$ is Gaussian in the
observables; here ${\bf \mu}$ is the vector of mean values of the observables, ${\bf
C}$ is their covariance matrix, and $_{,i}$ denotes a derivative with respect to
$i$th model parameter $p_i$. The parameter vector $\vec{p}$ includes 
both cosmological and any other model parameters needed to characterize the 
observations. This expression often simplifies --- for
example, for $N$ observable quantities with mean values ${\bf O}_{\alpha}$ and a
covariance matrix ${\bf C}$ that does not depend on the cosmological
parameters, the Fisher matrix becomes $F_{ij}=\sum_{\alpha, \beta} (\partial
{\bf O}_{\alpha}/\partial p_i) {\bf C}^{-1}_{\alpha\beta}(\partial {\bf
O}_{\beta}/\partial p_j)$.

By the Cramer-Rao inequality, a model parameter $p_i$ cannot be measured to a
precision better than $1/\sqrt{F_{ii}}$ when all other parameters are fixed, or
a precision $\sqrt{F^{-1}_{ii}}$ when all other parameters are marginalized
over.  In practice, the Fisher matrix is a good approximation to the
uncertainties as long as the likelihood can be approximated by a Gaussian,
which is generally the case near the peak of the likelihood and therefore in
cases when the parameters are measured with small errors.  Conversely, if the
errors are large, then the likelihood is typically non-Gaussian, and the
constraint region is no longer elliptical but characteristically banana-shaped,
as in Fig.~\ref{fig:allen06}. In this case, the Fisher matrix typically
underestimates the true parameter errors and degeneracies, and one should
employ a Monte Carlo approach to error estimation.

\bibliography{bibfile}

\end{document}